\documentclass[aps,prb,reprint,superscriptaddress,showpacs]{revtex4-2}

\usepackage{amsfonts}
\usepackage{amsmath}
\usepackage{amssymb}
\usepackage{graphicx}
\usepackage{dcolumn}
\usepackage{bm}
\usepackage{float}
\usepackage{hyperref}

\begin{document}

\title
  {
Spin selectivity  induced by non-collinear spins in Rashba wires
}

\author{Luciano Jacopo D'Onofrio}
\email{lucianojacopo.donofrio@spin.cnr.it}
\affiliation{CNR-SPIN, c/o Universit\`a di Salerno, IT-84084 Fisciano (SA), Italy}

\author{Maria Teresa Mercaldo}
\affiliation{Dipartimento di Fisica ``E. R. Caianiello", Universit\`a di Salerno, IT-84084 Fisciano (SA), Italy}

\author{Mario Cuoco}
\affiliation{CNR-SPIN, c/o Universit\`a di Salerno, IT-84084 Fisciano (SA), Italy}

\author{Carmine Ortix}
\email{cortix@unisa.it}
\affiliation{Dipartimento di Fisica ``E. R. Caianiello", Universit\`a di Salerno, IT-84084 Fisciano (SA), Italy}
\affiliation{CNR-SPIN, c/o Universit\`a di Salerno, IT-84084 Fisciano (SA), Italy}

\begin{abstract}
We report a general physical mechanism to obtain highly efficient spin selectivity in conventional
time-reversal symmetric one-dimensional systems without invoking phase decoherence or strong electronic correlations. We reveal that Rashba quantum wires featuring non-collinear spin states at the Fermi level inherently possess spin-selective transport properties. We show that this spin noncollinearity can be systematically designed and engineered by introducing an additional pseudospin degree of freedom---such as valley, sublattice, or orbital angular momentum---into spin-orbit coupled systems. By applying this framework to mirror-symmetric multi-subband semiconducting quantum wires and oxide nanowires, we establish a generalized route toward quantum-coherent spin selectivity up to $\sim 10 \%$. Our findings offer practical design principles for spin-selective transport devices without chirality.
\vspace{2ex}
\begin{center}
\includegraphics[width=0.75\columnwidth]{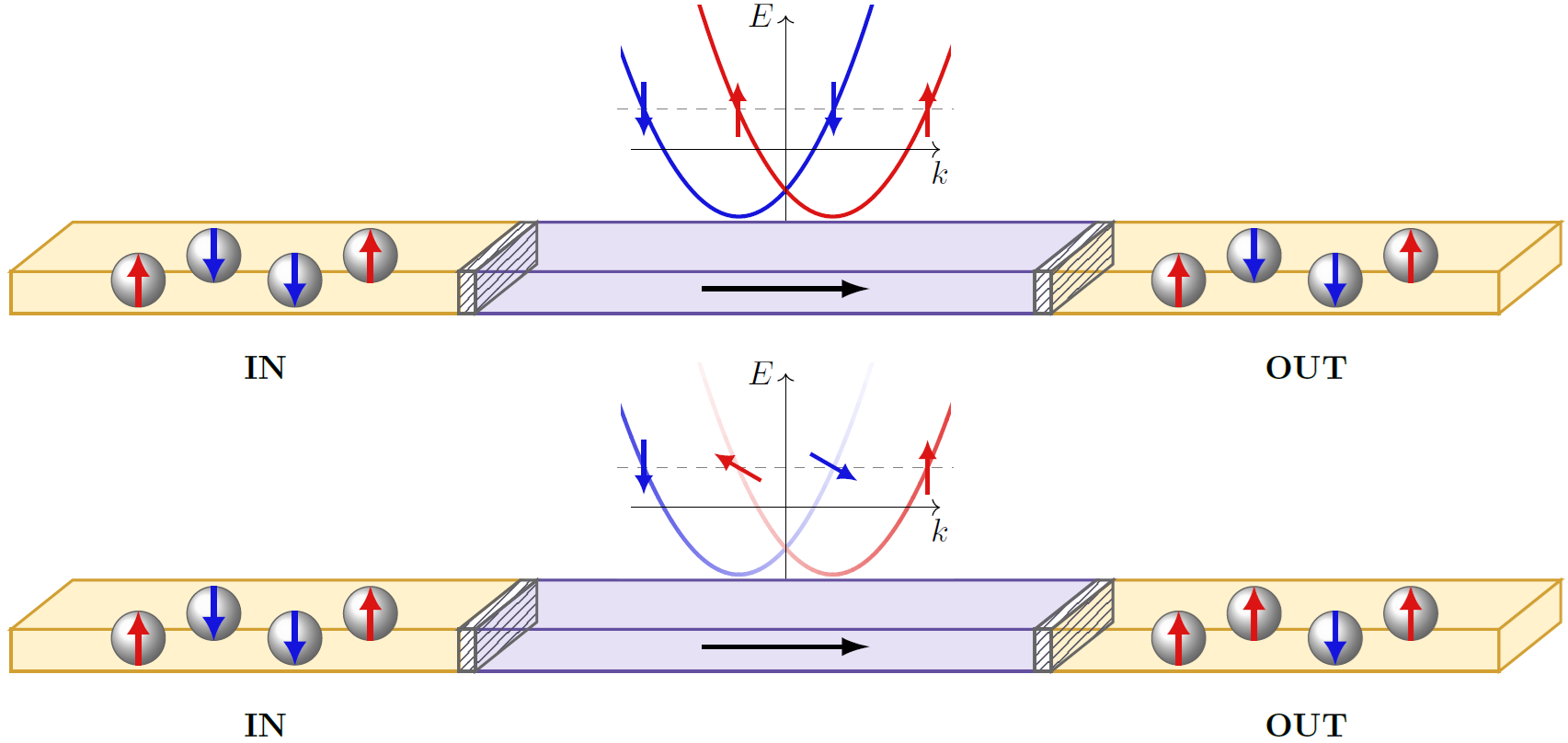}
\end{center}
\end{abstract}

\maketitle

The conversion between charge and spin
is a central concept in spintronics
since it enables efficient control of magnetism by voltage switches \cite{Matsukura2015,Manchon2019}. This efficient control can be achieved in materials with substantial spin-orbit coupling (SOC) using different mechanisms. In the spin Hall effect (SHE) \cite{Dyakonov1971,Hirsch1999,Valenzuela2006}, for instance, a longitudinal charge current generates a transversal spin current that results in a spin accumulation at surfaces. By contrast, the Edelstein effect \cite{Edelstein1990} corresponds to the non-equilibrium spin accumulation induced by charge currents in systems where inversion symmetry is broken\cite{Kato2004,Sanchez2013,Varotto2022,soumyanarayanan_nat16}. This spin accumulation is uniform in space and directly related to the momentum-dependent spin textures at the Fermi level. Moreover, spin-related conversion phenomena originating from orbital mechanisms can give rise to a wide range of effects, including the orbital Hall effect \cite{Tanaka2008,Kontani2009,Go2018,Cysne2022,Sala2023,Lyalin2023,Choi2023}, the orbital Edelstein effect \cite{Yoda_2018,Johansson2021, Go2017, Salemi2019, Chirolli2022}, the inverse orbital Edelstein effect \cite{Gaiardoni_2026, ElHamdi2023}, orbital photocurrents \cite{Adamantopoulos2024}, orbital filtering \cite{donofrio2025}, and orbital generation of Berry curvature \cite{les23, Mercaldo2023}.

In one-dimensional systems, the presence of spin-orbit coupling can also lead to spin selectivity.
Quantum mechanically, this occurs if the electron transmission probability through a material structure is different for opposite spins. This physical phenomenon
has recently attracted significant attention with the discovery of the Chiral-Induced Spin Selectivity (CISS) effect \cite{Ray1999-ex,Naaman2019}.
Originally observed in double-stranded DNA \cite{Gohler2011}, and later generalized to single-stranded DNA and other helical biological molecules \cite{Naaman2020}, the CISS has overcome the conventional wisdom according to which the simultaneous presence of strong spin-orbit coupling and magnetism is a necessary condition to have spin selectivity in one-dimensional systems \cite{Streda2003,Pershin2004,Debald2005}.
The presence of heavy elements with sizable atomic SOC has been suggested to be replaced by an effective spin-orbit coupling field due to the helical motion along the molecules. However, the observation of spin selectivity in the absence of time-reversal symmetry breaking poses a more fundamental question. This is evidenced by the fact that already the most paradigmatic model for transport in spin-orbit coupled one-dimensional systems --  a Rashba quantum wire -- is equipped with spin filtering capabilities only in the absence of  time-reversal symmetry\cite{Streda2003,Pershin2004,Debald2005}.
Recent theories have overcome this hurdle and suggested that a substantial spin selectivity can be achieved even in time-reversal symmetric conditions
when mechanisms of quantum mechanical dephasing \cite{Guo2012SpinSelective,Guo2014SpinDependent}, modelled for instance by B\"uttiker virtual leads \cite{Buttiker1986}, or strong electronic correlations~\cite{daghofer26,1denis26} are present. Spin selectivity with time-reversal symmetry has also been recently shown to occur in multiorbital helical junctions due to a complex-quaternion structure of the scattering matrix~\cite{UEA20}.

However,
the correspondence between chirality
and spin filtering capabilities is challenged when considering the physical properties of two-dimensional topological crystalline insulators, such as thin films of SnTe and Pb$_{(1-x)}$Sn$_{x}$Te. Their one-dimensional edge states, topologically protected by mirror Chern number \cite{Liu2014SpinFiltered},
are spin-filtered:
at each edge, there is an even number of right-moving conducting states which have the same mirror eigenvalue and hence spin polarization.
This implies that spin $\uparrow$ electrons coming from a lead will tunnel into the edge states with mirror eigenvalue $+i$ with 100\% certainty.
This capability raises the question of whether and how a quantum coherent spin filtering can occur in other non-topological one-dimensional systems
without chirality.
In this Letter, we will address precisely this question and show that a number of conventional one-dimensional systems, ranging from semiconducting quantum wires to one-dimensional oxide interfaces, can display a substantial spin filtering
even if they feature a mirror symmetry.
We will argue that Rashba wires with non-collinear spins of the quantum eigenstates at the Fermi level are naturally equipped with spin selectivity. Importantly, we will show that such spin noncollinearity can be designed in spin-orbit coupled systems with an additional pseudospin degree of freedom, such as valley, sublattice, or orbital angular momentum.  Two-subband semiconducting quantum wires and one-dimensional structures patterned at oxide interfaces are specific realizations of systems where spin noncollinearity, and consequently spin selectivity, can occur.

\begin{figure*}[t]
\centering
\includegraphics[width=0.75\textwidth]{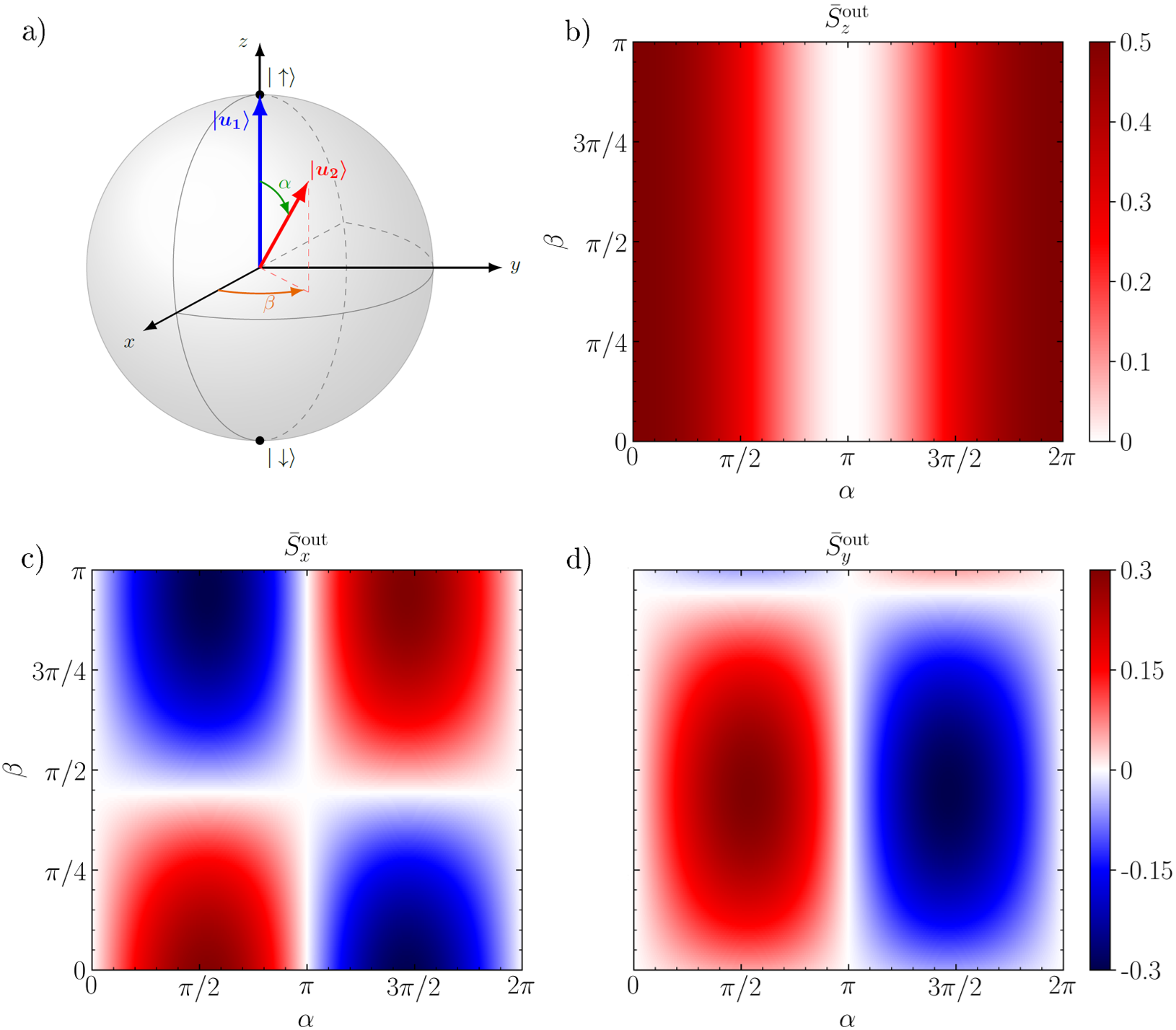}
\caption{Spin polarization acquired by unpolarized electrons transmitted through a Rashba wire hosting non-collinear spin states within the effective spin-$1/2$ single-channel approximation. (a) Bloch sphere representation of the spins of the right-moving states $| u_1 \rangle$ and $| u_2 \rangle$ at the Fermi level with $\alpha$ and $\beta$ the polar and azimuthal angle of the $| u_2 \rangle$ state. (b--d) Contour maps of the net spin polarization components in the metallic drain $\bar{S}_{z}^{\text{out}}$ (b), $\bar{S}_{x}^{\text{out}}$ (c), and $\bar{S}_{y}^{\text{out}}$ (d) acquired by unpolarized electrons as a function of $\alpha$ and $\beta$.}
\label{figure1}
\end{figure*}

Let us start by discussing how the presence of non-collinear spin states in one-dimensional systems can lead to spin-selective transport.
We assume that, precisely as in quantum wires with Rashba spin-orbit coupling, the electronic bands are spin split due to inversion symmetry breaking.
This directly implies that the four occupied states at the Fermi level can be grouped in two pairs,  each consisting of time-reversal symmetric partners. The spin expectation value and the velocity of the states making up a pair
will therefore be opposite. Time-reversal symmetry does not impose any constraint on the relative direction between the spins of states belonging to different pairs, which can therefore take any value. Consider now the relative angle between the spins of the two states with equal sign of the velocity (for instance the two right-moving states) that we dub $| u_1 \rangle$ and $| u_2\rangle$.
Parallel spins, {\it i.e.}, $\alpha \rightarrow 0$ [see Fig.~\ref{figure1}(a)], mimic the situation realized by the spin-filtered edge states of SnTe thin films. Antiparallel spins with $\alpha=\pi$ instead correspond to the situation realized in a single-mode quantum wire with Rashba spin-orbit coupling.
We will now
provide an argument to
show that for all values of the angle $\alpha \neq \pi$, spin selective transport naturally occurs.
Consider incoming spins entering the quantum wire, which propagate coherently along
only
two different channels  $| u_1\rangle$ and $| u_2\rangle$ with Fermi wavevectors $k_1$ and $k_2$, different due to the splitting of the bands.
We will assume in addition that the two propagation channels can be described by a two-component spinor.
When the spins leave the quantum wire,
the spinorial wavefunction
will therefore be a generic superposition
$\psi (x) = e^{ikx}\left[A|\uparrow\rangle + B|\downarrow\rangle \right]$ characterized by an outgoing spin polarization $S_{i}^{\text{out}} = \langle \psi|\hat{S}_{i}|\psi\rangle / \langle \psi|\psi\rangle$ with  $\hat{S}_i=\hat{\sigma}_i/2$, $\hat{\sigma}_i$ as the $i$-th Pauli matrix, and $i=x,y,z$.
Summing the direction of the incoming spin over the $4 \pi$ solid angle, we can directly compute [see the Supplemental Material] the net spin polarization acquired by unpolarized electrons. Fig.~\ref{figure1} (b),(c),(d) show the ensuing maps in the $\alpha-\beta$ plane with $\alpha$ the polar angle and $\beta$ the azimuthal angle. As expected for a single mode quantum wire with Rashba spin-orbit coupling, spin selective transport is precluded when the spins of the $| u_1\rangle$ and $| u_2\rangle$ states are antiparallel [see the Supplemental Material], {\it i.e.}, for $\alpha=\pi$. Conversely, for $\alpha \rightarrow 0$ electrons tend to be perfectly polarized with $\bar{S}_{z}^{\text{out}} \rightarrow 1/2$. Interestingly, we find that for generic values of $\alpha$, the outgoing spin polarization is always non-vanishing and, additionally, has a finite in-plane component except for specific values of the azimuthal angle $\beta$ [see Fig.~\ref{figure1}(c,d)]. This feature, however, is a result of an interference effect that sensitively depends on the
difference between the two momenta $k_1$ and $k_2$
and the length of the one-dimensional transmitting wire.

\begin{figure*}[t]
\centering
\includegraphics[width=0.75\textwidth]{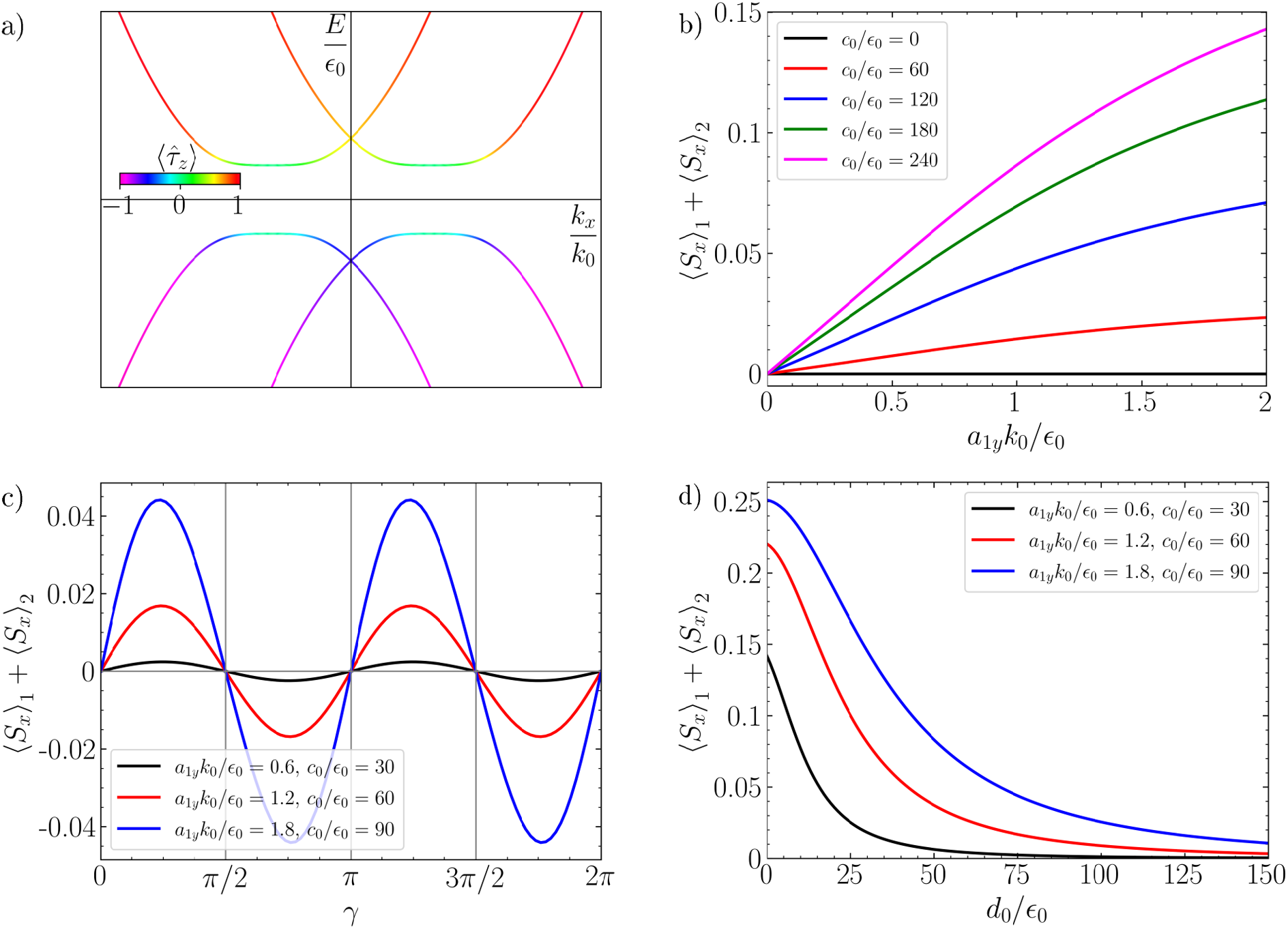}
\caption{Spin selectivity in an electron--hole wire with Rashba spin-orbit coupling. (a) Energy band dispersion
assuming model parameters $d_{2}k_{0}^{2}/\epsilon_{0}=1/200$, $a_{1y}k_{0}/\epsilon_{0}=1.8$, $c_{0}/\epsilon_{0}=90$, $\gamma=\pi/4$, and $d_{0}/\epsilon_{0}=35$.
We have introduced the characteristic energy scale $\epsilon_0$ and a reference momentum  $k_0$.
The color scale encodes the orbital character of the bands.
(b) Spin expectation value $S_x$ summed over the two right-moving states as a function of the intraband Rashba coupling strength $a_{1y}$ for different values of the interorbital mixing strength $c_0$. We have set the energy separation $d_0=70 \epsilon_0$ and the Fermi energy $\epsilon_F=10 \epsilon_0$. (c)-(d) Same as a function of the angle parameter $\gamma$ (c) and as a function of the energy separation $d_0$ (d).}
\label{figure2}
\end{figure*}

The argument provided above holds as an effective ``single-channel'' approximation [see the Supplemental Material]. In fact, in a conventional time-reversal invariant electronic system with spin as the only internal degree of freedom, such non-collinearity cannot be achieved in practice. We next show that the opposite is true in one-dimensional systems with multiple degrees of freedom.
To illustrate this,  we first consider a minimal four-band model that describes a conduction and a valence band, each of which contains a spin-orbit coupling term that is linear in momentum.
The resulting Hamiltonian can be written as ${\mathcal H}_0=d_0 {\hat S}_0 \otimes \hat{\tau}_z+ d_{2} k_{x}^{2} {\hat S}_0 \otimes \hat{\tau}_{z}+a_{1y}k_{x}\hat{S}_{y} \otimes {\hat \tau}_0 $,
where the $\hat{S}$'s matrices act in spin space, while the $\hat{\tau}$'s pseudospins
correspond to an orbital (or sublattice) degree of freedom.
The parameter $d_0$ controls the energy separation between the conduction and the valence band, while $d_2$ is related to the effective mass. Additionally, the parameter $a_{1y}$ is the strength of the intraband Rashba spin-orbit coupling.
Using the Hamiltonian above, one finds that,
setting the Fermi level position either in the conduction or in the valence band,
the spins of the two right-moving states are always antiparallel,
which precludes any spin-selective transport.
The situation changes dramatically by including spin-dependent interorbital couplings that are compatible with time-reversal symmetry. The total low-energy Hamiltonian then reads
${\mathcal{H}}={\mathcal H}_0+c_{0}(\cos\gamma\hat{S}_{x}+\sin\gamma\hat{S}_{y}) \otimes \hat{\tau}_{y}$,
with $c_0$ as the characteristic energy scale controlling the spin-dependent interorbital mixing. As shown in Fig.~\ref{figure2}(a), we find that at $c_0 \neq 0$ the bottom of the conduction band and the top of the valence bands possess a mixed orbital character, as also evidenced by a flattening of the bands.
Perhaps even more importantly, the spin-dependent interorbital mixing leads to a loss of collinearity between the spins of the two right-moving states. This can be measured by monitoring the behavior of the $\hat{x}$ component of the spin expectation value of the two right-moving states ($| u_1^+ \rangle$ and $| u_2^+ \rangle$) summed together.
A finite value of $\langle u_1^+| {\hat S}_x |  u_1^+\rangle + \langle u_2^+ | {\hat S}_x |  u_2^+ \rangle $ implies that the spins of the right-moving states rotate away from the $\hat{y}$ direction of the Rashba field but with a different angle.

This, in turn, guarantees their noncollinearity, and, in view of our foregoing discussion, a spin-selective transport. We find that this feature generally occurs for $c_0 \neq 0$ and the effect is enhanced by either increasing the strength of the intraorbital Rashba spin-orbit coupling or the interorbital mixing term [see Fig.~\ref{figure2}(b)].
We also find that the degree of noncollinearity strongly depends on the spin content of the interorbital mixing term. In fact, for $\gamma=n \pi/2$ with $n$ integer the spins of the right-moving states remain perfectly collinear and oriented along the Rashba field [see Fig.~\ref{figure2}(c)]. In particular, for $\gamma=\pi/2,3\pi/2$ the states remain spin eigenstates since the total Hamiltonian commutes with ${\hat S}_y \otimes {\hat \tau}_0$. Finally, we also find that when increasing the bare energy separation between the conduction and valence bands, governed by $d_0$ [see Fig.~\ref{figure2}(d)], we recover the complete absence of noncollinearity in a single-mode Rashba wire.
Altogether, these features suggest that spin selectivity can in principle be observed in narrow-gap semiconductor quantum wires with strong spin-orbit coupling.

\begin{figure*}[t]
\centering
\includegraphics[width=\textwidth]{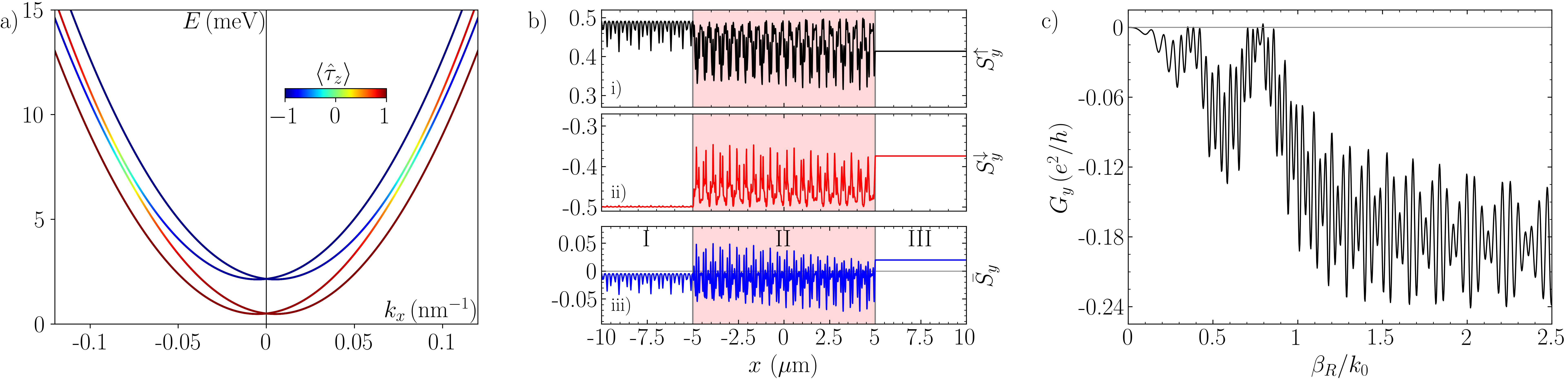}
\caption{
Spin selectivity in a two-subband InAs nanowire. (a) Energy dispersion for the model Hamiltonian in Eq.~\ref{eq:HamInAs}. (b) Spatial evolution of the spin $\hat{y}$ component injected from a two-subband metallic lead (region I) and collected in the outer metallic lead (region III). The scattering region II corresponds to the two-subband InAs nanowire. The top panel (i) corresponds to the injection of an $\uparrow$ spin, while the middle panel (ii) is for an $\downarrow$ spin. The net spin polarization resulting from the injection of unpolarized electrons, with $\rho_2/\rho_1=1$, is shown in the bottom panel (iii). (c) Spin conductance $G_y$ as a function of $\beta_R$ (in units of $k_0$) computed within the scattering-matrix formalism for unpolarized injection. The intersubband coupling, proportional to $\beta_R$, mixes the two subbands and makes the spins at the Fermi level non-collinear. For $\beta_R=0$ there is no spin imbalance at the Fermi level and $G_y$ vanishes. Spin conductance displays oscillations that average to $\sim0.18e^2/h$ for $\beta_R/k_0>1$. In all panels, we have set the nanowire length to 10 $\mu$m and the Fermi energy to $0.5$ meV above the energy of the Kramers doublet of the highest-energy subband. In addition, we have considered the model parameters $a=25 \epsilon_0 / k_0^2$, with $\epsilon_0=0.38$~meV and $k_0=0.1$nm$^{-1}$, $\alpha_R=2.6 \epsilon_0/k_0$, and $\Delta E=4.2 \epsilon_0$. In panels (a) and (b), $\beta_R=0.21 k_0$. The metallic leads have a parabolic dispersion $a^{\text{I},\text{III}} k_x^2$, with $a^{\text{I},\text{III}}=30 \epsilon_0 / k_0^2$; we have set the bottom of the bands at $0.51$~meV below the Kramers' doublet of the lowest-energy InAs subband.}
\label{figure3}
\end{figure*}

We next generalize this finding by showing that spin noncollinearity can be also achieved in a quasi-one-dimensional system
without chirality, featuring
two parabolic subbands that are spin split due to the Rashba spin-orbit coupling. Importantly,
a spin-dependent intersubband mixing similar in nature to the spin-dependent interorbital mixing discussed above naturally arises in this material platform by quantum confinement effects [see the Supplemental Material]. Specifically, confining in a transversal direction a two-dimensional electron gas with Rashba spin-orbit coupling leads to a two-subband one-dimensional Hamiltonian

\begin{align}
{\mathcal H}= a k_x^2 \hat{S}_0 \otimes \hat{\tau}_0 + 2 \alpha_R k_x {\hat S}_y \otimes \hat{\tau}_0 - 2 \alpha_R \beta_R  {\hat S}_x \otimes {\hat \tau}_y \nonumber \\ - \dfrac{\Delta E}{2} \hat{S}_0 \otimes \hat{\tau}_z + \textrm{const}
\label{eq:HamInAs}
\end{align}
where the $\hat{S}$'s matrices act in spin space and the $\hat{\tau}$'s Pauli matrices now act in subband space.
The Hamiltonian above possesses a vertical mirror symmetry that can be represented as ${\mathcal M_{x}}=- i \hat{S}_x \otimes \hat{\tau}_0$.
Fig.~\ref{figure3}(a) shows the resulting energy dispersion of the two subbands
considering the material dependent parameters $a,\alpha_R$ of InAs \cite{Stanescu2011} and a transversal size of the quasi-one-dimensional channel $d_y=130$~nm.
We find [see Fig.~\ref{figure3}(a)] that the parabolic behavior of the subbands is substantially altered at finite momenta with the presence of avoided level crossings where the electronic states have a mixed subband character. Close to these regions the four right-moving states (two per subband) lose their collinearity.
To verify that this characteristic implies spin selective transport, we have
first
solved the scattering problem for
single electrons
entering the two-subband wire from a spin-orbit free two-subband metallic lead [see the Supplemental Material]. Considering incoming spins that are parallel to the $\hat y$-oriented Rashba field,  we have [see Fig.~\ref{figure3}(b)] that the noncollinearity of the four right-moving states in the two-subband wire leads to a depolarization of the incoming spin both when parallel and when antiparallel to the Rashba field. However, this depolarization is different for the two configurations thus implying that incoming
unpolarized spins will develop a net spin polarization.
We find that the occurrence of this spin polarization is
immediately reflected in the occurrence of a finite spin conductance, which we compute by employing
the Landauer scattering matrix approach [see the Supplemental Material]. The intersubband coupling parametrized by $\beta_R$ generally leads to a finite spin conductivity [see Fig.~\ref{figure3}(c)] $G_y=\frac{e^2}{h} \mathrm{Tr}[t t^{\dagger} (\mathbb{I} \otimes \hat{\sigma}_y)]$, being $t$ the transmission matrix. We note that this finite spin conductivity does not contradict Bardarson's theorem~\cite{Bardarson_2008}, according to which the eigenvalues of $t t^{\dagger}$ must come in degenerate pairs.
The fact that our numerical calculations predict a spin
conductance that can reach values $\simeq 10 \%$ of the total unpolarized charge conductance $G=\frac{e^2}{h} \mathrm{Tr}[t t^{\dagger}]$
suggests that two-subband InAs nanowires can act as efficient spin filters.

\begin{figure*}[t]
\centering
\includegraphics[width=0.8\textwidth]{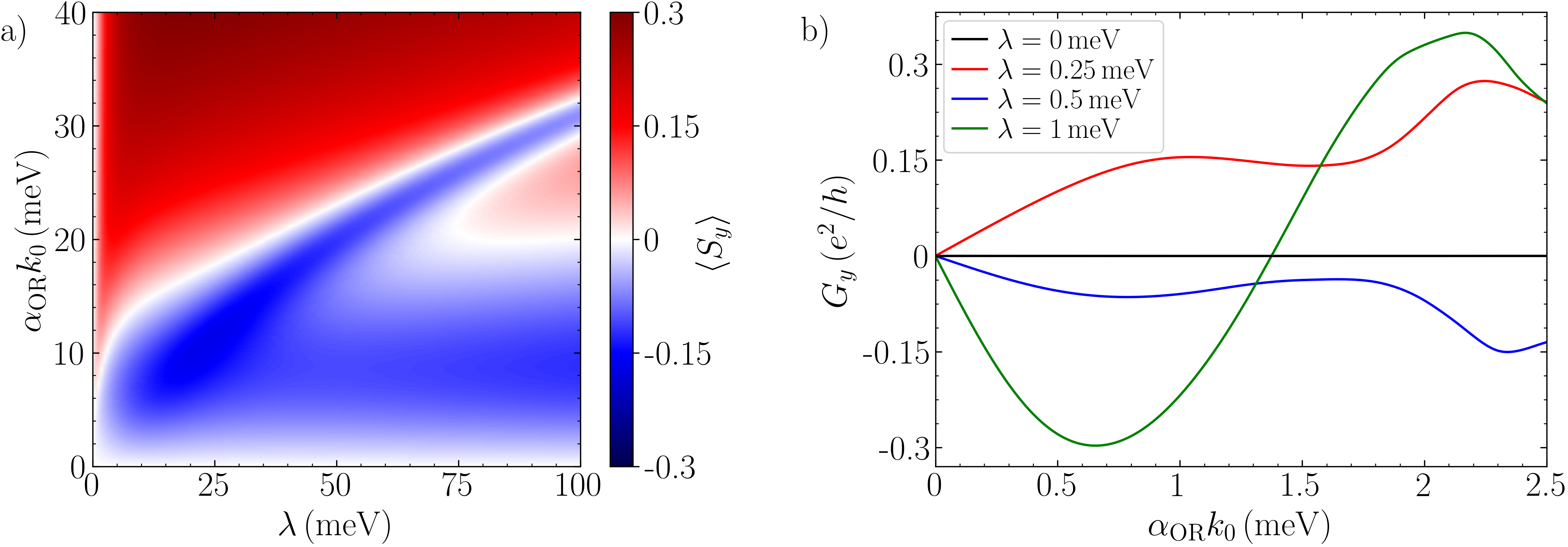}
\caption{Non-collinear spins and spin selectivity in an oxide nanowire with orbital Rashba and spin-orbit coupling. The nanowire is modelled by the Hamiltonian
${\mathcal H}=A k_x^2 + \alpha_{OR} k_x \hat{L}_y + \lambda \hat{{\bf L}} \cdot \hat{{\bf S}} + \delta_x \hat{L}_x^2 + \delta_y \hat{L}_y^2 + \delta_z \hat{L}_z^2$ with ${\hat L}_{x,y,z}$ the $L=1$ angular momentum operators. (a) Density plot of the ${\hat S}_y$ expectation value summed over the six right-moving states in the atomic spin-orbit coupling strength $\lambda$, orbital Rashba strength $\alpha_{OR}$ plane. We have set $A=0.15 \epsilon_0 / k_0^2$ with $\epsilon_0=38$meV and $k_0=1$~nm$^{-1}$, and  the crystal field parameters  $\delta_x=-0.2 \epsilon_0$, $\delta_y=-0.2 \epsilon_0$, $\delta_z=0.3 \epsilon_0$. As expected, $\langle S_y\rangle=0$ in the non-relativistic limit $\lambda\rightarrow0$.
(b) Spin conductance $G_y$ as a function of the orbital Rashba coupling $\alpha_{\mathrm{OR}}$ for different values of the atomic spin-orbit coupling $\lambda$, computed within the scattering-matrix formalism for unpolarized injection. $G_y$ vanishes identically for $\lambda=0$ and for $\alpha_{\mathrm{OR}}=0$.
To solve the scattering problem, we have considered two metallic leads with three parabolic subbands having the same effective mass coefficient as the oxide nanowire. In the source, we have also considered crystal field splittings with
$\delta_x^I=0.4 \epsilon_0$, $\delta_y^I=-0.5 \epsilon_0$ and $\delta_z^I=-0.7 \epsilon_0$. The scattering region length has been set to $d_x=1$~$\mu$m. The Fermi energy has been fixed at $2.5$~meV above the Kramers doublet of the upper orbital band.}
\label{figure4}
\end{figure*}

We finally predict that in the pure one-dimensional limit
spin noncollinearity can be designed in the absence of electron-hole mixing in a completely different material platform: complex oxide heterostructures. Interfaces of these materials host two-dimensional $d$ electron systems of $t_{2g}$ orbital character \cite{Popovic2008PRL,Salluzzo2009PRL,Khalsa2013PRB}, with one-dimensional nanowires that have been already synthesized \cite{Cen2008NatMater,Cen2009Science,Barthelemy2021EPL}. In these material structures, the degenerate $t_{2g}$ manifold spans an effective $L_{eff}=1$ angular momentum subspace. The absence of inversion symmetry that is naturally realized at the interfaces then has a twofold effect. First, there is a crystal field splitting between the orbitals transforming as $|p_x\rangle$, $|p_y\rangle$, $|p_z\rangle$, whose form depends on whether a tetragonal or orthorhombic phase is realized.
Second, there exists an interorbital mixing term which is linear in momentum and referred to as orbital Rashba coupling ~\cite{Park2011,park12, Khalsa2013PRB, kim14, Mercaldo2020}. This orbital Rashba coupling has been shown to cause a number of electronic and spintronic transport phenomena, ranging from the orbital Hall effect \cite{Go2018PRL, Go2021NatCommun} to a nonlinear Hall effect with time-reversal symmetry induced by the Berry curvature dipole \cite{Mercaldo2023,les23}. As we will discuss below, the orbital Rashba coupling is also at the origin of spin-selective transport. To show this, we first recall that
spin-orbit coupling effects can be relevant in such heterostructures, as for instance in LaAlO$_3$/SrTiO$_3$ heterostructures \cite{Caviglia2010PRL} and at KTaO$_3$ interfaces \cite{Ren2022SciAdv}. This motivates us to consider the orbital Rashba coupling, spin-orbit coupling and the crystal field on an equal footing. Assuming a single subband, the minimal low-energy theory is thus a six-band model,
which, precisely as the two subband InAs nanowire model, possesses a vertical mirror symmetry represented by ${\mathcal M}_x=- i  \hat{S}_x (1- 2 \hat{L}_x^2)$ with $\hat{S}_x$ and $\hat{L}_x$ the $\hat{x}$ component of spin and orbital angular momentum respectively. We determine the spin selectivity of this model by computing the spin conductance within the scattering matrix approach in a device where the oxide nanowire is attached
to two metallic leads which  have neither spin-orbit coupling nor orbital Rashba coupling.
To assess whether this model system can lead to spin noncollinearity, we first compute the spin content of the six right-moving states at the Fermi level.
Fig.~\ref{figure4}(a) shows the total $\hat{y}$ component of the spin expectation value.
The magnitude and the sign of the total spin can be tuned through the interplay between the orbital Rashba coupling and the atomic spin-orbit coupling $\lambda$. In particular, for any given value of $\lambda$ the increase of the orbital Rashba coupling leads to a reversal of the sign of $\langle S_y \rangle$ with a sizable amplitude that can be achieved both at small and large values of $\alpha_{OR}$. Remarkably, the net spin can remain sizable even in the regime of extremely weak atomic spin-orbit coupling $\lambda$. This behavior highlights the crucial role played by the orbital degrees of freedom and by the orbital Rashba coupling, whose interplay with the crystal-field potential can generate a pronounced spin noncollinearity independent of the presence of a strong spin-orbit coupling.
We thus focus on this regime and solve the scattering
matrix
to verify that the spin noncollinearity leads to spin selectivity. Fig.~\ref{figure4}(b) shows the
spin conductance $G_y$
as a function of the strength of the orbital Rashba coupling $\alpha_{OR}$ for values of the spin-orbit coupling smaller than $1$meV. The fact that we find
a spin conductance as large as $8\%$ the value of the unpolarized charge conductance
suggests that, as for two-subband InAs nanowires, also oxide nanowires can represent highly efficient spin filtering devices.

To wrap up, we have
unveiled a general physical
mechanism to obtain phase-coherent spin-selective transport in one-dimensional spin-orbit coupled systems. This phenomenon is entirely due to the presence of time-reversal related pairs at the Fermi level with spins that are non-collinear.
This non-collinearity is an intrinsic bulk property of the Rashba wire and thus offers a practical guide to identify systems with spin selectivity capabilities.
Its degree will clearly depend on the mesoscopic device properties including lead coupling, interface scattering, and the accumulated dynamical phases. We have shown that
spin noncollinearity, which is compatible with the presence of time-reversal symmetry,  generally occurs in systems with multiple internal degrees of freedom, including subbands and atomic orbitals. Our numerical calculations, using realistic material parameters, indicate that two-subband InAs Rashba quantum wires can develop a spin selectivity of $\sim 10\%$.
Oxide nanowires where the effects of orbital Rashba coupling, spin-orbit coupling, and crystal fields have been considered on an equal footing show comparable spin filtering efficiencies. Importantly, such efficiencies can be achieved even considering orbital Rashba coupling and spin-orbit coupling of a few meV.
Furthermore, the spin conductance displays oscillations as a function of the chemical potential with periods in the meV scale [see the Supplemental Material], thus implying that spin filtering effects can be observed up to 10 K.
Our predictions can be therefore tested in a variety of one-dimensional nanostructures.

\begin{acknowledgments}
M.C. and L.J.D'O. acknowledge partial support from
NRRP MUR project PE0000023-NQSTI. M.C. acknowledges support by Italian Ministry of University and Research (MUR) PRIN 2022 under the Grant No. 2022LP5K7 (BEAT).
\end{acknowledgments}

\bibliography{biblio}

@article{Bardarson_2008,
doi = {10.1088/1751-8113/41/40/405203},
url = {https://doi.org/10.1088/1751-8113/41/40/405203},
year = {2008},
month = {sep},
publisher = {},
volume = {41},
number = {40},
pages = {405203},
author = {Bardarson, J H},
title = {A proof of the Kramers degeneracy of transmission eigenvalues from antisymmetry of the scattering matrix},
journal = {Journal of Physics A: Mathematical and Theoretical}
}

@article{UEA20,
  title = {Spin selectivity through time-reversal symmetric helical junctions},
  author = {Utsumi, Yasuhiro and Entin-Wohlman, Ora and Aharony, Amnon},
  journal = {Phys. Rev. B},
  volume = {102},
  issue = {3},
  pages = {035445},
  numpages = {17},
  year = {2020},
  month = {Jul},
  publisher = {American Physical Society},
  doi = {10.1103/PhysRevB.102.035445},
  url = {https://link.aps.org/doi/10.1103/PhysRevB.102.035445}
}

@misc{daghofer26,
      title={Electronic correlations driving Chirality-Induced Spin Selectivity}, 
      author={Jacek Herbrych and Maria Daghofer},
      year={2026},
      eprint={2605.30240},
      archivePrefix={arXiv},
      primaryClass={cond-mat.str-el},
      url={https://arxiv.org/abs/2605.30240}, 
}

@misc{1denis26,
      title={Spin modulations in the Rashba-Hubbard chain -- a tensor network study}, 
      author={Jozef Genzor and Roman Krčmár and Andrej Gendiar and Chia-Min Chung and Denis Kochan},
      year={2026},
      eprint={2606.26014},
      archivePrefix={arXiv},
      primaryClass={cond-mat.str-el},
      url={https://arxiv.org/abs/2606.26014}, 
}

@article{Ren2022SciAdv,
  author  = {Ren, Tianshuang and Li, Miaocong and Sun, Xikang and Ju, Lele and Liu, Yuan and Hong, Siyuan and Sun, Yanqiu and Tao, Qian and Zhou, Yi and Xu, Zhu-An and Xie, Yanwu},
  title   = {Two-dimensional superconductivity at the surfaces of KTaO$_3$ gated with ionic liquid},
  journal = {Science Advances},
  volume  = {8},
  number  = {22},
  pages   = {eabn4273},
  year    = {2022},
  doi     = {10.1126/sciadv.abn4273}
}

@article{Caviglia2010PRL,
  author  = {Caviglia, Andrea D. and Gabay, Marcelo and Gariglio, Stefano and
             Reyren, Nicolas and Cancellieri, Claudia and Triscone, Jean-Marc},
  title   = {Tunable Rashba Spin-Orbit Interaction at Oxide Interfaces},
  journal = {Physical Review Letters},
  volume  = {104},
  pages   = {126803},
  year    = {2010},
  doi     = {10.1103/PhysRevLett.104.126803}
}

@article{Go2021NatCommun,
  author  = {Go, Dongwook and Jo, Daegeun and Gao, Tong and Ando, Kazuya and Bl{\"u}gel, Stefan and Manchon, Aur{\'e}lien and Lee, Hyun-Woo},
  title   = {Orbital Rashba effect in a surface-oxidized Cu thin film},
  journal = {Nature Communications},
  volume  = {12},
  pages   = {4964},
  year    = {2021},
  doi     = {10.1038/s41467-021-25292-1}
}

@article{Go2018PRL,
  author  = {Go, Dongwook and Jo, Daegeun and Kim, Changyoung and Lee, Hyun-Woo},
  title   = {Intrinsic Spin and Orbital Hall Effects from Orbital-Dependent Band Geometric Properties},
  journal = {Physical Review Letters},
  volume  = {121},
  pages   = {086602},
  year    = {2018},
  doi     = {10.1103/PhysRevLett.121.086602}
}

@article{Salluzzo2009PRL,
  author  = {Salluzzo, M. and Cezar, J. C. and Brookes, N. B. and Bisogni, V. and De Luca, G. M. and Richter, C. and Thiel, S. and Mannhart, J. and Huijben, M. and Brinkman, A. and Ghiringhelli, G.},
  title   = {Orbital Reconstruction and the Two-Dimensional Electron Gas at the LaAlO$_3$/SrTiO$_3$ Interface},
  journal = {Physical Review Letters},
  volume  = {102},
  pages   = {166804},
  year    = {2009},
  doi     = {10.1103/PhysRevLett.102.166804}
}

@article{Popovic2008PRL,
  author  = {Popović, Z. S. and Satpathy, S. and Martin, R. M.},
  title   = {Origin of the Two-Dimensional Electron Gas Carrier Density at the LaAlO$_3$ on SrTiO$_3$ Interface},
  journal = {Physical Review Letters},
  volume  = {101},
  pages   = {256801},
  year    = {2008},
  doi     = {10.1103/PhysRevLett.101.256801}
}

@article{Barthelemy2021EPL,
  author  = {Barthelemy, A. and Bergeal, N. and Bibes, M. and
             Caviglia, A. D. and Citro, R. and Cuoco, M. and
             Kalaboukhov, A. and Kalisky, B. and Perroni, C. A. and
             Santamaria, J. and Stornaiuolo, D. and Salluzzo, M.},
  title   = {Quasi-two-dimensional electron gas at the oxide interfaces for topological quantum physics},
  journal = {EPL (Europhysics Letters)},
  volume  = {133},
  number  = {1},
  pages   = {17001},
  year    = {2021},
  doi     = {10.1209/0295-5075/133/17001}
}

@article{Cen2009Science,
  author    = {Cen, Cheng and Thiel, Stefan and Mannhart, Jochen and Levy, Jeremy},
  title     = {Oxide Nanoelectronics on Demand},
  journal   = {Science},
  volume    = {323},
  number    = {5917},
  pages     = {1026--1030},
  year      = {2009},
  doi       = {10.1126/science.1168294},
  publisher = {American Association for the Advancement of Science}
}

@ARTICLE{Khalsa2013PRB,
  title = {Theory of ${t}_{2g}$ electron-gas Rashba interactions},
  author = {Khalsa, Guru and Lee, Byounghak and MacDonald, A. H.},
  journal = {Phys. Rev. B},
  volume = {88},
  issue = {4},
  pages = {041302},
  numpages = {5},
  year = {2013},
  month = {Jul},
  publisher = {American Physical Society},
  doi = {10.1103/PhysRevB.88.041302},
  url = {https://link.aps.org/doi/10.1103/PhysRevB.88.041302}
}

@article{kim14,
author = {Kim, P. and Kang, K. T. and Go, G. and Han, J. H.}, 
journal = {Phys. Rev. B},
title={Nature of orbital and spin Rashba coupling in the surface bands of {SrTiO$_3$} and {KTaO$_3$}},
volume = {90}, 
pages = {205423},
year = {2014}
}

@ARTICLE{Mercaldo2020,
  title = {Electrically Tunable Superconductivity Through Surface Orbital Polarization},
  author = {Mercaldo, Maria Teresa and Solinas, Paolo and Giazotto, Francesco and Cuoco, Mario},
  journal = {Phys. Rev. Applied},
  volume = {14},
  issue = {3},
  pages = {034041},
  numpages = {15},
  year = {2020},
  month = {Sep},
  publisher = {American Physical Society},
  doi = {10.1103/PhysRevApplied.14.034041},
  url = {https://link.aps.org/doi/10.1103/PhysRevApplied.14.034041}
}

@article{park12,
author = {Park, J.-H. and Kim, C. H. and Rhim, J.-W. and Han, J. H.}, 
journal = {Phys. Rev. B},
title={Orbital {Rashba} effect and its detection by circular dichroism angle-resolved photoemission spectroscopy},
volume = {85},
pages = {195401},
year = {2012}
}

@article{Park2011,
  title = {Orbital-Angular-Momentum Based Origin of {Rashba} -Type Surface Band Splitting},
  author = {Park, Seung Ryong and Kim, Choong H. and Yu, Jaejun and Han, Jung Hoon and Kim, Changyoung},
  journal = {Phys. Rev. Lett.},
  volume = {107},
  issue = {15},
  pages = {156803},
  numpages = {5},
  year = {2011},
  month = {Oct},
  publisher = {American Physical Society},
  doi = {10.1103/PhysRevLett.107.156803},
  url = {https://link.aps.org/doi/10.1103/PhysRevLett.107.156803}
}

@article{Cen2008NatMater,
  author    = {Cen, Cheng and Thiel, Stefan and Hammerl, German and
               Schneider, Christof W. and Andersen, K. E. and
               Hellberg, C. Stephen and Mannhart, Jochen and Levy, Jeremy},
  title     = {Nanoscale Control of an Interfacial Metal--Insulator Transition at Room Temperature},
  journal   = {Nature Materials},
  volume    = {7},
  number    = {4},
  pages     = {298--302},
  year      = {2008},
  doi       = {10.1038/nmat2136},
  publisher = {Nature Publishing Group}
}

@article{donofrio2025,
  title = {Filtering spin and orbital moment in centrosymmetric systems},
  author = {D'Onofrio, Luciano Jacopo and Mercaldo, Maria Teresa and Brzezicki, Wojciech and K\l{}osi\ifmmode \acute{n}\else \'{n}\fi{}ski, Adam and Mazzola, Federico and Ortix, Carmine and Cuoco, Mario},
  journal = {Phys. Rev. B},
  volume = {112},
  issue = {8},
  pages = {085428},
  numpages = {13},
  year = {2025},
  month = {Aug},
  publisher = {American Physical Society},
  doi = {10.1103/2th8-nyz8},
  url = {https://link.aps.org/doi/10.1103/2th8-nyz8}
}

@article{Stanescu2011,
  title = {Majorana fermions in semiconductor nanowires},
  author = {Stanescu, Tudor D. and Lutchyn, Roman M. and Das Sarma, S.},
  journal = {Phys. Rev. B},
  volume = {84},
  issue = {14},
  pages = {144522},
  numpages = {29},
  year = {2011},
  month = {Oct},
  publisher = {American Physical Society},
  doi = {10.1103/PhysRevB.84.144522},
  url = {https://link.aps.org/doi/10.1103/PhysRevB.84.144522}
}

@article{Gohler2011,
  author  = {G{\"o}hler, Benjamin and Hamelbeck, Volker and Markus, T. Z. 
             and Kettner, Michael and Hanne, G. F. and Vager, Zeev 
             and Naaman, Ron and Zacharias, Helmut},
  title   = {Spin Selectivity in Electron Transmission Through 
             Self-Assembled Monolayers of Double-Stranded DNA},
  journal = {Science},
  volume  = {331},
  pages   = {894--897},
  year    = {2011},
  doi     = {10.1126/science.1199339}
}

@article{Naaman2020,
  author  = {Naaman, Ron and Paltiel, Yossi and Waldeck, David H.},
  title   = {Chiral Molecules and the Spin Selectivity Effect},
  journal = {The Journal of Physical Chemistry Letters},
  volume  = {11},
  pages   = {3660--3666},
  year    = {2020},
  doi     = {10.1021/acs.jpclett.0c00474}
}

@article{Matsukura2015,
  author  = {Matsukura, Fumihiro and Tokura, Yoshinori and Ohno, Hideo},
  title   = {Control of magnetism by electric fields},
  journal = {Nature Nanotechnology},
  volume  = {10},
  pages   = {209--220},
  year    = {2015},
  doi     = {10.1038/nnano.2015.22}
}

@article{Manchon2019,
  author  = {Manchon, A. and Zelezny, J. and Miron, I. M. and Jungwirth, T.
             and Sinova, J. and Thiaville, A. and Garello, K. and Gambardella, P.},
  title   = {Current-induced spin-orbit torques in ferromagnetic and antiferromagnetic systems},
  journal = {Reviews of Modern Physics},
  volume  = {91},
  pages   = {035004},
  year    = {2019},
  doi     = {10.1103/RevModPhys.91.035004}
}

@article{Liu2014SpinFiltered,
  author  = {Liu, Junwei and Hsieh, Timothy H. and Wei, Peng and Duan, Wenhui and Moodera, Jagadeesh S. and Fu, Liang},
  title   = {Spin-filtered edge states with an electrically tunable gap in a two-dimensional topological crystalline insulator},
  journal = {Nature Materials},
  volume  = {13},
  pages   = {178--183},
  year    = {2014},
  doi     = {10.1038/nmat3828}
}

@article{Edelstein1990,
  author  = {V. M. Edelstein},
  title   = {Spin polarization of conduction electrons induced by electric current in two-dimensional asymmetric electron systems},
  journal = {Solid State Communications},
  volume  = {73},
  number  = {3},
  pages   = {233--235},
  year    = {1990},
  doi     = {10.1016/0038-1098(90)90963-C}
}

@article{Debald2005,
  author    = {S. Debald and B. Kramer},
  title     = {Rashba effect and magnetic field in semiconductor quantum wires},
  journal   = {Physical Review B},
  volume    = {71},
  pages     = {115322},
  year      = {2005},
  doi       = {10.1103/PhysRevB.71.115322},
  publisher = {American Physical Society}
}

@article{Pershin2004,
  author    = {Yuriy V. Pershin and James A. Nesteroff and Vladimir Privman},
  title     = {Effect of spin-orbit interaction and in-plane magnetic field on the conductance of a quasi-one-dimensional system},
  journal   = {Physical Review B},
  volume    = {69},
  pages     = {121306(R)},
  year      = {2004},
  doi       = {10.1103/PhysRevB.69.121306},
  publisher = {American Physical Society}
}

@article{Kato2004,
  author  = {Y. K. Kato and R. C. Myers and A. C. Gossard and D. D. Awschalom},
  title   = {Current-Induced Spin Polarization in Strained Semiconductors},
  journal = {Physical Review Letters},
  volume  = {93},
  pages   = {176601},
  year    = {2004},
  doi     = {10.1103/PhysRevLett.93.176601},
  publisher = {American Physical Society}
}

@article{Guo2012SpinSelective,
  author  = {Ai-Min Guo and Qing-feng Sun},
  title   = {Spin-Selective Transport of Electrons in DNA Double Helix},
  journal = {Physical Review Letters},
  volume  = {108},
  pages   = {218102},
  year    = {2012},
  doi     = {10.1103/PhysRevLett.108.218102}
}

@article{Guo2014SpinDependent,
  author  = {Ai-Min Guo and Qing-Feng Sun},
  title   = {Spin-dependent electron transport in protein-like single-helical molecules},
  journal = {Proceedings of the National Academy of Sciences of the United States of America},
  volume  = {111},
  number  = {32},
  pages   = {11658--11662},
  year    = {2014},
  doi     = {10.1073/pnas.1407716111}
}

@article{Buttiker1986,
  title = {Role of quantum coherence in series resistors},
  author = {B\"uttiker, M.},
  journal = {Phys. Rev. B},
  volume = {33},
  issue = {5},
  pages = {3020--3026},
  numpages = {0},
  year = {1986},
  month = {Mar},
  publisher = {American Physical Society},
  doi = {10.1103/PhysRevB.33.3020},
  url = {https://link.aps.org/doi/10.1103/PhysRevB.33.3020}
}

@article{Streda2003,
  title = {Antisymmetric Spin Filtering in One-Dimensional Electron Systems with Uniform Spin-Orbit Coupling},
  author = {Streda, P. and  Seba, P.},
  journal = {Phys. Rev. Lett.},
  volume = {90},
  issue = {25},
  pages = {256601},
  numpages = {4},
  year = {2003},
  month = {Jun},
  publisher = {American Physical Society},
  doi = {10.1103/PhysRevLett.90.256601},
  url = {https://link.aps.org/doi/10.1103/PhysRevLett.90.256601}
}

@article{Cysne2022,
  title = {Orbital {Hall} effect in bilayer transition metal dichalcogenides: From the intra-atomic approximation to the Bloch states orbital magnetic moment approach},
  author = {Cysne, Tarik P. and Bhowal, Sayantika and Vignale, Giovanni and Rappoport, Tatiana G.},
  journal = {Phys. Rev. B},
  volume = {105},
  issue = {19},
  pages = {195421},
  numpages = {15},
  year = {2022},
  month = {May},
  publisher = {American Physical Society},
  doi = {10.1103/PhysRevB.105.195421},
  url = {https://link.aps.org/doi/10.1103/PhysRevB.105.195421}
}

@article{Adamantopoulos2024,
  title = {Orbital {Rashba} Effect as a Platform for Robust Orbital Photocurrents},
  author = {Adamantopoulos, T. and Merte, M. and Go, D. and Freimuth, F. and Bl\"ugel, S. and Mokrousov, Y.},
  journal = {Phys. Rev. Lett.},
  volume = {132},
  issue = {7},
  pages = {076901},
  numpages = {7},
  year = {2024},
  month = {Feb},
  publisher = {American Physical Society},
  doi = {10.1103/PhysRevLett.132.076901},
  url = {https://link.aps.org/doi/10.1103/PhysRevLett.132.076901}
}

@article{les23,
	Author = {Lesne, Edouard and Sa{\v g}lam, Yildiz G. and Battilomo, Raffaele and Mercaldo, Maria Teresa and van Thiel, Thierry C. and Filippozzi, Ulderico and Noce, Canio and Cuoco, Mario and Steele, Gary A. and Ortix, Carmine and Caviglia, Andrea D.},
	Da = {2023/05/01},
	Doi = {10.1038/s41563-023-01498-0},
	Id = {Lesne2023},
	Isbn = {1476-4660},
	Journal = {Nature Materials},
	Number = {5},
	Pages = {576--582},
	Title = {{Designing spin and orbital sources of Berry curvature at oxide interfaces}},
	Ty = {JOUR},
	Url = {https://doi.org/10.1038/s41563-023-01498-0},
	Volume = {22},
	Year = {2023}}

@Article{Salemi2019,
author={Salemi, Leandro
and Berritta, Marco
and Nandy, Ashis K.
and Oppeneer, Peter M.},
title={Orbitally dominated {Rashba-Edelstein} effect in noncentrosymmetric antiferromagnets},
journal={Nature Communications},
year={2019},
month={Nov},
day={26},
volume={10},
number={1},
pages={5381},
issn={2041-1723},
doi={10.1038/s41467-019-13367-z},
url={https://doi.org/10.1038/s41467-019-13367-z}
}

@Article{Go2017,
author={Go, Dongwook
and Hanke, Jan-Philipp
and Buhl, Patrick M.
and Freimuth, Frank
and Bihlmayer, Gustav
and Lee, Hyun-Woo
and Mokrousov, Yuriy
and Bl\"ugel, Stefan},
title={Toward surface orbitronics: giant orbital magnetism from the orbital {Rashba} effect at the surface of sp-metals},
journal={Scientific Reports},
year={2017},
month={Apr},
day={25},
volume={7},
number={1},
pages={46742},
issn={2045-2322},
doi={10.1038/srep46742},
url={https://doi.org/10.1038/srep46742}
}

@article{Yoda_2018,
   title={Orbital Edelstein Effect as a Condensed-Matter Analog of Solenoids},
   volume={18},
   ISSN={1530-6992},
   url={http://dx.doi.org/10.1021/acs.nanolett.7b04300},
   DOI={10.1021/acs.nanolett.7b04300},
   number={2},
   journal={Nano Letters},
   publisher={American Chemical Society (ACS)},
   author={Yoda, Taiki and Yokoyama, Takehito and Murakami, Shuichi},
   year={2018},
   month=feb, pages={916–920} }

@article{Johansson2021,
  title = {Spin and orbital Edelstein effects in a two-dimensional electron gas: Theory and application to {{SrTiO}}$_{3}$ interfaces},
  author = {Johansson, Annika and G\"obel, B\"orge and Henk, J\"urgen and Bibes, Manuel and Mertig, Ingrid},
  journal = {Phys. Rev. Res.},
  volume = {3},
  issue = {1},
  pages = {013275},
  numpages = {9},
  year = {2021},
  month = {Mar},
  publisher = {American Physical Society},
  doi = {10.1103/PhysRevResearch.3.013275},
  url = {https://link.aps.org/doi/10.1103/PhysRevResearch.3.013275}
}

@Article{Varotto2022,
author={Varotto, Sara
and Johansson, Annika
and G\"obel, B\"orge
and Vicente-Arche, Luis M.
and Mallik, Srijani
and Br\'ehin, Julien
and Salazar, Rapha\'el
and Bertran, Francois
and F\'evre, Patrick Le
and Bergeal, Nicolas
and Rault, Julien
and Mertig, Ingrid
and Bibes, Manuel},
title={Direct visualization of {Rashba}-split bands and spin/orbital-charge interconversion at {KTaO$_3$} interfaces},
journal={Nature Communications},
year={2022},
month={Oct},
day={18},
volume={13},
number={1},
pages={6165},
issn={2041-1723},
doi={10.1038/s41467-022-33621-1},
url={https://doi.org/10.1038/s41467-022-33621-1}
}

@Article{Naaman2019,
author={Naaman, Ron
and Paltiel, Yossi
and Waldeck, David H.},
title={Chiral molecules and the electron spin},
journal={Nature Reviews Chemistry},
year={2019},
month={Apr},
day={01},
volume={3},
number={4},
pages={250-260},
issn={2397-3358},
doi={10.1038/s41570-019-0087-1},
url={https://doi.org/10.1038/s41570-019-0087-1}
}

@ARTICLE{Ray1999-ex,
  title    = "Asymmetric scattering of polarized electrons by organized organic films of chiral molecules",
  author   = "Ray, K and Ananthavel, S P and Waldeck, D H and Naaman, R",
  journal  = "Science",
  volume   =  283,
  number   =  5403,
  pages    = "814--816",
  month    =  feb,
  year     =  1999
}

@article{soumyanarayanan_nat16,
  title={Emergent phenomena induced by spin--orbit coupling at surfaces and interfaces},
  author={Soumyanarayanan, Anjan and Reyren, Nicolas and Fert, Albert and Panagopoulos, Christos},
  journal={Nature},
  volume={539},
  number={7630},
  pages={509--517},
  year={2016},
  publisher={Nature Publishing Group UK London}
}

@Article{Sanchez2013,
author={S{\'a}nchez, J. C. Rojas
and Vila, L.
and Desfonds, G.
and Gambarelli, S.
and Attan{\'e}, J. P.
and De Teresa, J. M.
and Mag{\'e}n, C.
and Fert, A.},
title={Spin-to-charge conversion using {Rashba} coupling at the interface between non-magnetic materials},
journal={Nature Communications},
year={2013},
month={Dec},
day={17},
volume={4},
number={1},
pages={2944},
issn={2041-1723},
doi={10.1038/ncomms3944},
url={https://doi.org/10.1038/ncomms3944}
}

@article{Tanaka2008,
  title = {Intrinsic spin {Hall} effect and orbital {Hall} effect in $4d$ and $5d$ transition metals},
  author = {Tanaka, T. and Kontani, H. and Naito, M. and Naito, T. and Hirashima, D. S. and Yamada, K. and Inoue, J.},
  journal = {Phys. Rev. B},
  volume = {77},
  issue = {16},
  pages = {165117},
  numpages = {16},
  year = {2008},
  month = {Apr},
  publisher = {American Physical Society},
  doi = {10.1103/PhysRevB.77.165117},
  url = {https://link.aps.org/doi/10.1103/PhysRevB.77.165117}
}

@ARTICLE{Dyakonov1971,
       author = {{D'Yakonov}, M.~I. and {Perel'}, V.~I.},
        title = "{Possibility of orienting electron spins with current}",
      journal = {Soviet Journal of Experimental and Theoretical Physics Letters},
         year = 1971,
        month = jun,
       volume = {13},
        pages = {467},
       adsurl = {https://ui.adsabs.harvard.edu/abs/1971JETPL..13..467D}
}

@Article{Valenzuela2006,
author={Valenzuela, S. O.
and Tinkham, M.},
title={Direct electronic measurement of the spin {Hall} effect},
journal={Nature},
year={2006},
month={Jul},
day={01},
volume={442},
number={7099},
pages={176-179},
issn={1476-4687},
doi={10.1038/nature04937},
url={https://doi.org/10.1038/nature04937}
}

@article{Hirsch1999,
  title = {Spin {Hall} Effect},
  author = {Hirsch, J. E.},
  journal = {Phys. Rev. Lett.},
  volume = {83},
  issue = {9},
  pages = {1834--1837},
  numpages = {0},
  year = {1999},
  month = {Aug},
  publisher = {American Physical Society},
  doi = {10.1103/PhysRevLett.83.1834},
  url = {https://link.aps.org/doi/10.1103/PhysRevLett.83.1834}
}

@article{Kontani2009,
  title = {Giant Orbital {Hall} Effect in Transition Metals: Origin of Large Spin and Anomalous {Hall} Effects},
  author = {Kontani, H. and Tanaka, T. and Hirashima, D. S. and Yamada, K. and Inoue, J.},
  journal = {Phys. Rev. Lett.},
  volume = {102},
  issue = {1},
  pages = {016601},
  numpages = {4},
  year = {2009},
  month = {Jan},
  publisher = {American Physical Society},
  doi = {10.1103/PhysRevLett.102.016601},
  url = {https://link.aps.org/doi/10.1103/PhysRevLett.102.016601}
}

@article{Go2018,
  title = {Intrinsic Spin and Orbital {Hall} Effects from Orbital Texture},
  author = {Go, Dongwook and Jo, Daegeun and Kim, Changyoung and Lee, Hyun-Woo},
  journal = {Phys. Rev. Lett.},
  volume = {121},
  issue = {8},
  pages = {086602},
  numpages = {6},
  year = {2018},
  month = {Aug},
  publisher = {American Physical Society},
  doi = {10.1103/PhysRevLett.121.086602},
  url = {https://link.aps.org/doi/10.1103/PhysRevLett.121.086602}
}

@article{Sala2023,
  title = {Orbital Hanle Magnetoresistance in a $3d$ Transition Metal},
  author = {Sala, Giacomo and Wang, Hanchen and Legrand, William and Gambardella, Pietro},
  journal = {Phys. Rev. Lett.},
  volume = {131},
  issue = {15},
  pages = {156703},
  numpages = {6},
  year = {2023},
  month = {Oct},
  publisher = {American Physical Society},
  doi = {10.1103/PhysRevLett.131.156703},
  url = {https://link.aps.org/doi/10.1103/PhysRevLett.131.156703}
}

@article{Chirolli2022,
  title = {Colossal Orbital {Edelstein} Effect in Noncentrosymmetric Superconductors},
  author = {Chirolli, Luca and Mercaldo, Maria Teresa and Guarcello, Claudio and Giazotto, Francesco and Cuoco, Mario},
  journal = {Phys. Rev. Lett.},
  volume = {128},
  issue = {21},
  pages = {217703},
  numpages = {6},
  year = {2022},
  month = {May},
  publisher = {American Physical Society},
  doi = {10.1103/PhysRevLett.128.217703},
  url = {https://link.aps.org/doi/10.1103/PhysRevLett.128.217703}
}

@Article{ElHamdi2023,
author={El Hamdi, Anas
and Chauleau, Jean-Yves
and Boselli, Margherita
and Thibault, Clementine
and Gorini, Cosimo
and Smogunov, Alexander
and Barreteau, Cyrille
and Gariglio, Stefano
and Triscone, Jean-Marc
and Viret, Michel},
title={Observation of the orbital inverse {Rashba-Edelstein} effect},
journal={Nature Physics},
year={2023},
month={Dec},
day={01},
volume={19},
number={12},
pages={1855-1860},
issn={1745-2481},
doi={10.1038/s41567-023-02121-4},
url={https://doi.org/10.1038/s41567-023-02121-4}
}

@Article{Mercaldo2023,
author={Mercaldo, Maria Teresa
and Noce, Canio
and Caviglia, Andrea D.
and Cuoco, Mario
and Ortix, Carmine},
title={Orbital design of {Berry} curvature: pinch points and giant dipoles induced by crystal fields},
journal={npj Quantum Materials},
year={2023},
month={Feb},
day={28},
volume={8},
number={1},
pages={12},
issn={2397-4648},
doi={10.1038/s41535-023-00545-y},
url={https://doi.org/10.1038/s41535-023-00545-y}
}

@article{Lyalin2023,
  title = {Magneto-Optical Detection of the Orbital {Hall} Effect in Chromium},
  author = {Lyalin, Igor and Alikhah, Sanaz and Berritta, Marco and Oppeneer, Peter M. and Kawakami, Roland K.},
  journal = {Phys. Rev. Lett.},
  volume = {131},
  issue = {15},
  pages = {156702},
  numpages = {6},
  year = {2023},
  month = {Oct},
  publisher = {American Physical Society},
  doi = {10.1103/PhysRevLett.131.156702},
  url = {https://link.aps.org/doi/10.1103/PhysRevLett.131.156702}
}

@Article{Choi2023,
author={Choi, Young-Gwan
and Jo, Daegeun
and Ko, Kyung-Hun
and Go, Dongwook
and Kim, Kyung-Han
and Park, Hee Gyum
and Kim, Changyoung
and Min, Byoung-Chul
and Choi, Gyung-Min
and Lee, Hyun-Woo},
title={Observation of the orbital {Hall} effect in a light metal {Ti}},
journal={Nature},
year={2023},
month={Jul},
day={01},
volume={619},
number={7968},
pages={52-56},
issn={1476-4687},
doi={10.1038/s41586-023-06101-9},
url={https://doi.org/10.1038/s41586-023-06101-9}
}

@article{Gaiardoni_2026,
  title = {Boltzmann theory of the inverse Edelstein effect in a two-dimensional Rashba gas},
  author = {Gaiardoni, Irene and Trama, Mattia and Maiellaro, Alfonso and Guarcello, Claudio and Romeo, Francesco and Citro, Roberta},
  journal = {Phys. Rev. B},
  volume = {113},
  issue = {19},
  pages = {195419},
  numpages = {11},
  year = {2026},
  month = {May},
  publisher = {American Physical Society},
  doi = {10.1103/8srn-y73s},
  url = {https://link.aps.org/doi/10.1103/8srn-y73s}
}

\clearpage
\onecolumngrid

\setcounter{equation}{0}
\setcounter{figure}{0}
\setcounter{table}{0}
\renewcommand{\theequation}{S\arabic{equation}}
\renewcommand{\thefigure}{S\arabic{figure}}
\renewcommand{\thetable}{S\arabic{table}}

\begin{center}
{\large\bfseries Supplemental Material:\\[2pt]
Spin selectivity  induced by non-collinear spins in Rashba wires}\\[10pt]
Luciano Jacopo D'Onofrio, Maria Teresa Mercaldo, Mario Cuoco, and Carmine Ortix
\end{center}
\vspace{1.5em}

\section*{Spin-split bands and spin imbalance: the effective spin-$1/2$ single-channel approximation}

To start, we present the effective spin-$1/2$ single-channel approximation. We consider a transmission system featuring a single propagation direction ($x$) and comprising three distinct regions (I, II, and III). We assume spin-degenerate electronic states with the same momentum $k$ at the Fermi level in the outer regions (I and III). Consequently, the generic initial electronic state injected from region I can be expressed as:
\begin{equation}
\psi^{\text{I}}(x) = e^{ikx}\left[\cos\dfrac{\theta}{2}|\uparrow\rangle + e^{i\phi}\sin\dfrac{\theta}{2}|\downarrow\rangle \right],
\end{equation}
where $\theta\in[0,\pi]$, $\phi\in[0,2\pi)$, and $|\uparrow\rangle, |\downarrow\rangle$ are the eigenstates of the Pauli $z$-matrix $\hat{\sigma}_{z}$. The corresponding transmitted state in region III is defined as:
\begin{equation}
\psi^{\text{III}}(x) = e^{ikx}\left[A^{\text{III}}|\uparrow\rangle + B^{\text{III}}|\downarrow\rangle \right].
\end{equation}

In the intermediate region II, spanning from $x=0$ to $x=d$, we suppose that the spin states at the Fermi level are non-collinear, namely:
\begin{equation}
\begin{cases}
| u_1\rangle = |\uparrow\rangle \\
| u_2\rangle = \cos\left(\dfrac{\alpha}{2}\right)|\uparrow\rangle + e^{i\beta}\sin\left(\dfrac{\alpha}{2}\right)|\downarrow\rangle
\end{cases},
\end{equation}
where $\alpha\in(0,2\pi)$ and $\beta\in[0,\pi]$, associated with electronic momenta $k_{1}$ and $k_{2}$, respectively. In doing so, we consider a situation in which there are only two available propagation channels with non-collinear spins
living in a two-dimensional Hilbert space.
On the other hand, as long as time-reversal symmetry is preserved, a non-collinear spin configuration at the Fermi level cannot be realized in an electronic system assuming only the internal spin degree of freedom. Spin non-collinearity requires indeed the presence of additional internal (orbital, sub-band or sublattice) degrees of freedom directly coupled to the spin, as shown in the main text. Our argument can be seen as an effective approximation in a Hilbert subspace without an underlying Hamiltonian and consequently well-defined transmission properties. Therefore, to evaluate the spin polarization in the outer region, we assume that the electrons enter and leave the intermediate region unimpeded, {\it i.e.}, we assume vanishing reflection amplitudes at the interfaces.

The state in the intermediate region is then constructed as
\begin{equation}
\begin{aligned}
\psi^{\text{II}}(x) &= A^{\text{II}}e^{ik_{1}x}| u_1\rangle + B^{\text{II}}e^{ik_{2}x}| u_2\rangle \\
 &= \left[A^{\text{II}}e^{ik_{1}x}+B^{\text{II}}e^{ik_{2}x}\cos\left(\dfrac{\alpha}{2}\right)\right]|\uparrow\rangle \\
 &\quad + B^{\text{II}}e^{i(\beta+k_{2}x)}\sin\left(\dfrac{\alpha}{2}\right)|\downarrow\rangle
\end{aligned}
\end{equation}

and the
following boundary conditions at the interfaces are consequently imposed:
\begin{equation}
\begin{cases}
|\psi^{\text{I}}(0)\rangle = |\psi^{\text{II}}(0)\rangle \\
|\psi^{\text{II}}(d)\rangle = |\psi^{\text{III}}(d)\rangle
\end{cases}.
\end{equation}
Solving this system for a given initial configuration $\left(\theta,\phi\right)$ yields the outgoing amplitudes:
\begin{equation}
\begin{cases}
A^{\text{III}} = \cos\dfrac{\theta}{2}K_{1} + e^{i(\phi-\beta)}\sin\dfrac{\theta}{2}\cot\left(\dfrac{\alpha}{2}\right)(K_{2}-K_{1})\\
B^{\text{III}} = K_{2}e^{i\phi}\sin\dfrac{\theta}{2}
\end{cases},
\end{equation}
with $K_{1,2} = e^{i(k_{1,2}-k)d}$. Having $\hat{S}_{i}= \hbar \hat{\sigma}_{i}/2$, the spin moment components for the outgoing state are then computed as $S_{i}^{\text{out}} = \langle \psi^{\text{III}}|\hat{S}_{i}|\psi^{\text{III}}\rangle / \langle \psi^{\text{III}}|\psi^{\text{III}}\rangle$ for $i=x,y,z$. This leads to
\begin{equation}
\begin{cases}
S_{x}^{\text{out}}(\theta,\phi,\alpha,\beta) = \dfrac{\Re\{ A^{\text{III},*}B^{\text{III}}\} }{|A^{\text{III}}|^{2}+|B^{\text{III}}|^{2}}\\
S_{y}^{\text{out}}(\theta,\phi,\alpha,\beta) = \dfrac{\Re\{ -iA^{\text{III},*}B^{\text{III}}\} }{|A^{\text{III}}|^{2}+|B^{\text{III}}|^{2}}\\
S_{z}^{\text{out}}(\theta,\phi,\alpha,\beta) = \dfrac{|A^{\text{III}}|^{2}-|B^{\text{III}}|^{2}}{2(|A^{\text{III}}|^{2}+|B^{\text{III}}|^{2})}
\end{cases}.
\end{equation}

To determine the average outgoing spin moment, we integrate over all possible initial configurations across the solid angle $\Omega$. Apart from accidental points dependent on the geometry of the system, the average outgoing spin vanishes when $\alpha=\pi$. In this specific configuration, the two states at the Fermi level in region II are $| u_1\rangle = |\uparrow\rangle$ and $| u_2\rangle = e^{i\beta}|\downarrow\rangle$, indicating no spin imbalance for the transmitting states. Substituting $\alpha=\pi$ provides
\begin{equation}
\begin{cases}
S_{x}^{\text{out}}(\theta,\phi,\pi,\beta) = \dfrac{1}{2}\cos[(k_{2}-k_{1})d+\phi]\sin\theta\\
S_{y}^{\text{out}}(\theta,\phi,\pi,\beta) = \dfrac{1}{2}\sin[(k_{2}-k_{1})d+\phi]\sin\theta\\
S_{z}^{\text{out}}(\theta,\phi,\pi,\beta) = \dfrac{1}{2}\cos\theta
\end{cases}.
\end{equation}
Integrating these components over the solid angle $\Omega$,
\begin{equation}
\bar{S}_{i}^{\text{out}}(\pi,\beta) = \dfrac{1}{4\pi}\int_{0}^{\pi}d\theta\sin\theta\int_{0}^{2\pi}d\phi\, S_{i}^{\text{out}}(\theta,\phi,\pi,\beta),
\end{equation}
confirms a vanishing transferred average spin moment ($\bar{S}_{x}^{\text{out}} = \bar{S}_{y}^{\text{out}} = \bar{S}_{z}^{\text{out}} = 0$).

\section*{Spin non-collinearity by coupling the spin to a pseudospin degree of freedom}

Let us consider a 1D system preserving time-reversal symmetry (TRS), characterized by a spin degree of freedom $S=1/2$ coupled to a pseudospin $\tau=1/2$ representing an internal degree of freedom (for instance, a doublet of energy subbands or orbitals).
This pseudospin is described by the Pauli matrices $\hat{\tau}_{i}$, with $i=x,y,z$ and only $\hat{\tau}_{y}$ odd under TRS.

Since the spin operator $\hat{\boldsymbol{S}}$ is odd under TRS, the product $\hat{\tau}_{i}\hat{S}_{j}$ breaks TRS for $i=x,z$, but preserves it for $i=y$. Given that the momentum operator $\hat{p}_{x}$ also changes sign under TRS, a
low-energy continuum
Hamiltonian that preserves this symmetry can be constructed by retaining only linear terms in the pseudospin and spin operators:
\begin{equation}
\begin{aligned}
{\mathcal{H}} &= \hat{p}_{x}^{2}\left[a_{2} + b_{2}\hat{\tau}_{x} + (\boldsymbol{c}_{2}\cdot\hat{\boldsymbol{S}})\hat{\tau}_{y} + d_{2}\hat{\tau}_{z}\right] \\
&\quad + \hat{p}_{x}\left[\boldsymbol{a}_{1}\cdot\hat{\boldsymbol{S}} + (\boldsymbol{b}_{1}\cdot\hat{\boldsymbol{S}})\hat{\tau}_{x} + c_{1}\hat{\tau}_{y} + (\boldsymbol{d}_{1}\cdot\hat{\boldsymbol{S}})\hat{\tau}_{z}\right] \\
&\quad + a_{0} + b_{0}\hat{\tau}_{x} + (\boldsymbol{c}_{0}\cdot\hat{\boldsymbol{S}})\hat{\tau}_{y} + d_{0}\hat{\tau}_{z},
\end{aligned}
\end{equation}
where the sets of coefficients $\left(a, b,c,d\right)$ parameterize the system.

Moving to the momentum space and fixing a specific parameter set, we can diagonalize ${\mathcal{H}}$ to obtain the band structure $E(k)$. For a given chemical potential $\mu$, we identify forward-propagating normalized eigenstates $|u_{n}^+\rangle$, defined by a positive group velocity ($\hbar^{-1} \partial E/\partial k > 0$).
The net spin at the Fermi level
is evaluated by summing the expectation value of that spin component over these transmitting states, i.e.,
\begin{equation}
 \sum_{n}\langle u_{n}^+|\hat{S}_{i}|u_{n}^+\rangle.
\end{equation}

\section*{InAs nanowires: a two subband model}

We consider electron transmission across  InAs nanowires, and we
focus on a quasi-1D regime with multi-subband occupancy. We assume strong confinement along the $z$-direction, such that only the lowest sub-band is occupied, while the weaker confinement along $y$ allows for multiple occupied sub-bands ($d_z \ll d_y \ll d_x$). Thus, by limiting our analysis to energy scales of a few meV, we construct a low-energy effective model.

We start from a tight-binding model on a simple cubic lattice with nearest-neighbor hopping. In the long-wavelength (continuous) limit ($k \to 0$), the system is described by a Hamiltonian that preserves time-reversal symmetry and includes the Rashba spin-orbit interaction (SOI) in the $xy$-plane:
\begin{equation}
{H} = a\hat{p}_{x}^{2} + b\hat{p}_{y}^{2} + 2\alpha_{R}\left(\hat{p}_{x}\hat{S}_{y} - \hat{p}_{y}\hat{S}_{x}\right),
\end{equation}
where $a$ and $b$ are related to the effective electron mass, and $\alpha_R$ is the Rashba spin-orbit coupling. We define a complete basis of states $|\Psi_{n}(x,y)\rangle = \psi_{n_{1}}(x)\phi_{n_{2}}(y)|\eta_{n_{3}}\rangle$, where $|\eta_{n_{3}}\rangle$ denotes the spin degree of freedom. Imposing hard-wall boundary conditions along the transverse direction ($|\Psi(x, y=\pm d_{y}/2)\rangle = 0$) yields the transverse states
\begin{equation}
\begin{cases}
\phi_{n}^{+}(y) =
\sqrt{\dfrac{2}{d_y}}\cos\left[\dfrac{(2n-1)\pi}{d_{y}}y\right]\\
\phi_{n}^{-}(y) =
\sqrt{\dfrac{2}{d_y}} \sin\left(\dfrac{2n\pi}{d_{y}}y\right)
\end{cases},
\end{equation}
with $n=1,2,\ldots$ . Applying the kinetic operator $b\hat{p}_{y}^{2}$ gives the corresponding energies $E_{y,n}^{+} = b(2n-1)^{2}\pi^{2}/d_{y}^{2}$ and $E_{y,n}^{-} = 4bn^{2}\pi^{2}/d_{y}^{2}$. Consequently, the two lowest-energy states, associated with the first two sub-bands, are $ \phi_{1}^{+} \rightarrow | \phi_1 \rangle$ and $ \phi_{1}^{-} \rightarrow | \phi_2 \rangle$, with energies $E_{1,y} = b\pi^{2}/d_{y}^{2}$ and $E_{2,y} = 4b\pi^{2}/d_{y}^{2}$, respectively.
The resulting
energy difference between the $k_x=0$ Kramers doublet of the first
two subbands is
given by $\Delta E = E_{2,y} - E_{1,y} = 3b\pi^{2}/d_{y}^{2}$.

To isolate the relevant low-energy dynamics, we project the Hamiltonian into the subspace spanned by these first two sub-bands:
\begin{equation}
\hat{h} = \begin{pmatrix}
\langle \phi_{1}|\hat{H}|\phi_{1}\rangle  & \langle \phi_{1}|\hat{H}|\phi_{2}\rangle \\
\langle \phi_{2}|\hat{H}|\phi_{1}\rangle  & \langle \phi_{2}|\hat{H}|\phi_{2}\rangle
\end{pmatrix} = \begin{pmatrix}
\hat{H}_{1,1} & \hat{H}_{1,2}\\
\hat{H}_{2,1} & \hat{H}_{2,2}
\end{pmatrix}.
\end{equation}
The inter-subband coupling is dictated by the matrix element $\langle \phi_{1}|\hat{p}_{y}|\phi_{2}\rangle = -i\beta_R$, where $\beta_R = \langle \phi_{1}|\partial_{y}|\phi_{2}\rangle = 8/(3d_{y})$. Therefore, the projected matrix elements become
\begin{equation}
\hat{h} = \begin{pmatrix}
a\hat{p}_{x}^{2} + E_{1,y} + 2\alpha_{R}\hat{p}_{x}\hat{S}_{y} & 2i\alpha_{R}\beta_R\hat{S}_{x}\\
-2i\alpha_{R}\beta_R\hat{S}_{x} & a\hat{p}_{x}^{2} + E_{2,y} + 2\alpha_{R}\hat{p}_{x}\hat{S}_{y}
\end{pmatrix}.
\end{equation}
Then, we introduce the pseudospin Pauli matrices $\hat{\tau}_{i}$ to represent the sub-band degree of freedom, with $|\phi_{1}\rangle = |\uparrow_{\tau}\rangle$ and $|\phi_{2}\rangle = |\downarrow_{\tau}\rangle$. Consequently, the effective Hamiltonian in momentum space takes the following compact form:
\begin{equation}
\hat{h}(k_x) = ak_x^{2} + 2\alpha_{R}k_x\hat{S}_{y} - 2\alpha_{R}\beta_R\hat{S}_{x}\hat{\tau}_{y} + \left(\dfrac{E_{1,y}-E_{2,y}}{2}\right)\hat{\tau}_{z} + \dfrac{E_{1,y}+E_{2,y}}{2}.
\label{eq:HamInAs-SM}
\end{equation}

Assuming an effective mass $m^{*} = \nu m_{0}$ (with $m_0$ the bare electron mass), the quadratic coefficient $a$ corresponds to $\hbar^{2}/(2\nu m_0)$. Furthermore, the sub-band energies can be expressed in terms of
$\Delta E$ as $E_{1,y} = \Delta E/3$ and $E_{2,y} = 4\Delta E/3$. For the specific case of an InAs nanowire, we adopt the following realistic material parameters: $\nu = 0.04$, $\Delta E = 1.6$ meV, $d_{y} = 130$ nm, and $\alpha_{R} = 10$ meV$\cdot$nm.

We aim to assess the electron transmission across a heterostructure made up of three regions: a source
lead (I), the InAs nanowire (II), and the drain
lead (III). The leads are modeled by the free-electron Hamiltonian
${\mathcal{H}}^{\text{A}}(k_x) = a^{\text{A}}k_x^{2}$,
with $\text{A}=\text{I, III}$. In these regions, for a given chemical potential $\mu$, the longitudinal wavevectors are
$k_{i}^{\text{A}} = \sqrt{\mu/a^{\text{A}}}$, with $i=1,2$.
On the other hand, region II is described by the Hamiltonian introduced in Eq.~\ref{eq:HamInAs-SM}: ${\mathcal{H}}^{\text{II}}(k_x) = \hat{h}(k_x)$, substituting $a \rightarrow a^{\text{II}}$.

For numerical implementation, we
put the Hamiltonians in dimensionless form
by dividing them by a characteristic energy scale $\epsilon_{0} = \hbar^{2}k_{0}^{2}/(2m_{0}) = \hbar^{2}/(2m_{0}d_{0}^{2})$ (with $k_{0}d_{0}=1$).
In our case, we select a length scale $d_{0} = 10$ nm and, as a consequence, the energy unit is $\epsilon_{0} \approx 0.38$ meV. In light of the realistic parameters mentioned above for the InAs nanowire, we obtain $\bar{a}^{\text{II}} = {a^{\text{II}}k_{0}^{2}}/{\epsilon_{0}} = 25$, $\bar{\alpha}_{R} = \alpha_Rk_0/ \varepsilon_0 =  2.6$, $\bar{\beta}_{R} = \beta_R / k_0 = 0.21$, $\bar{E}_{1,y}^{\text{II}} = {E}_{1,y}^{\text{II}} / \varepsilon_0 =  1.4$, and $\bar{E}_{2,y}^{\text{II}} = {E}_{2,y}^{\text{II}} / \varepsilon_0 =  5.6$. Moreover, we set $\bar{a}^{\text{I}} = \bar{a}^{\text{III}} = 30$, corresponding to
Fermi wavelength of the order of $100$ nm in the
leads for the chosen chemical potential.

\section*{One-dimensional $L=1$ model with orbital Rashba and spin-orbit couplings}

We study a one-dimensional system characterized by broken inversion symmetry, an internal orbital degree of freedom $L=1$ (e.g., $p$- or $t_{2g}$-orbitals), and an intrinsic atomic spin-orbit coupling (SOC). Similar to the previous model, we assume a strong transverse confinement along the $y$-direction ($d_y \ll d_x$). Under this geometrical constraint, the energy spacing between the transverse sub-bands, which scales as $d_y^{-2}$, is assumed to be significantly larger than the crystal field splitting
among the $L=1$ orbitals.
This energy hierarchy allows us to project the low-energy dynamics exclusively onto the lowest transverse sub-band ($E_{1,y}=0$), while retaining the three orbital degrees of freedom spanned by the states $|p_z\rangle$, $|p_y\rangle$, and $|p_x\rangle$.

Assuming the basis vectors are aligned with the reference axes, we adopt the standard representation for the atomic orbitals:
\begin{equation}
|p_z\rangle = \begin{pmatrix} 1 \\ 0 \\ 0 \end{pmatrix}, \quad |p_y\rangle = \begin{pmatrix} 0 \\ 1 \\ 0 \end{pmatrix}, \quad |p_x\rangle = \begin{pmatrix} 0 \\ 0 \\ 1 \end{pmatrix}.
\end{equation}
In this manifold, the components of the orbital angular momentum operator $\hat{\boldsymbol{L}}$ are defined as
\begin{equation}
\hat{L}_x = \begin{pmatrix} 0 & -i & 0 \\ i & 0 & 0 \\ 0 & 0 & 0 \end{pmatrix}, \quad
\hat{L}_y = \begin{pmatrix} 0 & 0 & -i \\ 0 & 0 & 0 \\ i & 0 & 0 \end{pmatrix}, \quad
\hat{L}_z = \begin{pmatrix} 0 & 0 & 0 \\ 0 & 0 & -i \\ 0 & i & 0 \end{pmatrix},
\end{equation}
satisfying the commutation relations $[\hat{L}_\alpha, \hat{L}_\beta] = i\epsilon_{\alpha \beta \gamma}\hat{L}_\gamma$.

The scattering system comprises a spin-orbit active central region (II), spanning from $x=-d_x/2$ to $x= d_x/2$, coupled, as before, to a source and a drain
lead (I and III).
The effective Hamiltonian in momentum space for a generic region ($\text{A=I, II, III}$) can be written as
\begin{equation}
{\mathcal{H}}^{\text{A}}(k_x) = \hat{A}^{\text{A}}k_{x}^{2} + \delta_{\text{A,II}}\left[\alpha_{\text{OR}}k_{x}(\cos\gamma\hat{L}_{y}-\sin\gamma\hat{L}_{z}) + \lambda\hat{\boldsymbol{L}}\cdot\hat{\boldsymbol{S}}\right] + \hat{\Delta}^{\text{A}},
\end{equation}
where $\hat{A}^{\text{A}} = \sum_{\alpha} a_{\alpha}^{\text{A}} \hat{L}_\alpha^2$ and $\hat{\Delta}^{\text{A}} = \sum_{\alpha} \delta_\alpha^{\text{A}} \hat{L}_\alpha^2$, with $\alpha =x,y,z$. The Kronecker delta $\delta_{\text{A,II}}$ activates the orbital Rashba coupling (with strength $\alpha_{\text{OR}}$ and orientation $\gamma$) and the SOC (with intensity $\lambda$) exclusively in region II, where the spatial inversion symmetry is broken.
In this region, the spin polarization $\langle S_i \rangle$ at the Fermi level can be evaluated for each spin component ($i=x,y,z$) by summing the expectation value of the corresponding spin operator $\hat{S}_i$ over the available forward-propagating states.

For the numerical simulations, we cast the Hamiltonian into a dimensionless form. We select a length unit $d_0 = 1$ nm, yielding a characteristic energy scale $\epsilon_0 \approx 38$ meV. To reproduce an effective electron mass $m^* \approx 3m_0$, we set $\bar{a}_{\alpha}^{\text{A}} = {a_{\alpha}^{\text{A}}k_{0}^{2}}/{\epsilon_{0}} = 0.15$ across all regions for $\alpha=x,y,z$. The dimensionless crystal field parameters, defined as $\bar{\delta}_\alpha^{\text{A}} = \delta_\alpha^{\text{A}}/\epsilon_0$, are chosen as $\bar{\delta}_x^{\text{I}} = 0.4$, $\bar{\delta}_y^{\text{I}} = -0.5$, and $\bar{\delta}_z^{\text{I}} = -0.7$ for the injection region (I), $\bar{\delta}_x^{\text{II}} = -0.2$, $\bar{\delta}_y^{\text{II}} = -0.2$, and $\bar{\delta}_z^{\text{II}} = 0.3$ for the intermediate region (II), and $\bar{\delta}_{\alpha}^{\text{III}} = 0$ for the collection region (III). In all simulations, the region II is assumed to have a length $d_x = 1\ \mu$m, the orientation angle $\gamma$ of the orbital Rashba interaction is 0.
Finally, we consider a wide parameter space by varying the spin-orbit coupling $\lambda$ between 0 and 100 meV,
and the orbital Rashba parameter
$\alpha_{\text{OR}}$ between 0 and 40 meV$\cdot$nm.

\section*{Single-electron scattering problem}

We start by noticing that for both the InAs and the $L=1$ nanowire models,
the Hamiltonian in both the scattering region and the leads ($\text{A}=\text{I},\text{II},\text{III}$) can be written in the symmetrized form
\begin{equation}
\mathcal{H}^{\text{A}}=\hat{p}_x \hat{A}^{\text{A}}\hat{p}_x+\tfrac{1}{2}\{\hat{p}_x,\hat{B}^{\text{A}}\}+\hat{C}^{\text{A}}.
\end{equation}
The operators $\hat{A}^{\text{A}}$, $\hat{B}^{\text{A}}$, and $\hat{C}^{\text{A}}$ contain the inverse effective masses, the terms linear in momentum (spin or orbital Rashba couplings), and the momentum-independent terms (subband or crystal-field splittings, intersubband and atomic spin-orbit couplings), respectively. In the leads, spin-orbit coupling is absent: $\hat{B}^{\text{A}}=0$, and $\hat{A}^{\text{A}}$ and $\hat{C}^{\text{A}}$ are spin independent.

The chemical potential $\mu$ is chosen such that each region hosts $N$ right-moving ($+$) and $N$ left-moving ($-$) propagating modes ($N=4$ and $N=6$ in the InAs and multiorbital nanowires, respectively), with wavevectors $k_n^{\text{A},\pm}$ and normalized eigenvectors $|u_n^{\text{A},\pm}\rangle$ defined by $\mathcal{H}^{\text{A}}(k_n^{\text{A},\pm})|u_n^{\text{A},\pm}\rangle=\mu|u_n^{\text{A},\pm}\rangle$, where $\mathcal{H}^{\text{A}}(k)=\hat{A}^{\text{A}}k^{2}+\hat{B}^{\text{A}}k+\hat{C}^{\text{A}}$. Since the Hamiltonian of the leads is spin-independent, we choose their modes as products $|\chi_s\rangle\otimes|\phi_m\rangle$ of a spin state $|\chi_s\rangle$ ($s=\uparrow,\downarrow$), eigenstate of $\hat{\sigma}_i$, and a subband (orbital) state $|\phi_m\rangle$ ($m=1,\ldots,N/2$).

We then consider a single electron injected from region I in a given spin and subband (orbital) state. In each region, the electronic state is expanded over the propagating modes as
\begin{equation}
|\psi^{\text{A}}(x)\rangle=\sum_{n=1}^{N}\left[c_{n}^{\text{A},+}\exp(ik_{n}^{\text{A},+}x)|u_{n}^{\text{A},+}\rangle+c_{n}^{\text{A},-}\exp(ik_{n}^{\text{A},-}x)|u_{n}^{\text{A},-}\rangle\right],
\end{equation}
where $c_n^{\text{III},-}=0$, since region III is assumed to contain only outgoing right-moving waves. For an electron injected in the channel $(m,s)$, the incoming wave is
\begin{equation}
|\psi_{\text{in}}^{m,s}(x)\rangle=\exp(ik_m^{\text{I}}x)|\chi_s\rangle\otimes|\phi_m\rangle,
\end{equation}
where $k_m^{\text{I}}$ is the wavevector of the right-moving mode $|\chi_s\rangle\otimes|\phi_m\rangle$ in region I. This wave fixes the coefficients $c_n^{\text{I},+}$, while the remaining $4N$ coefficients ($c_n^{\text{I},-}$, $c_n^{\text{II},\pm}$, and $c_n^{\text{III},+}$) are determined by the boundary conditions at the two interfaces, located at $x=-d_x/2$ (left) and $x=d_x/2$ (right). Integrating the Schr\"odinger equation across an interface shows that $|\psi\rangle$ and $|\tilde{\psi}\rangle=\hat{A}|\psi'\rangle+\tfrac{i}{2}\hat{B}|\psi\rangle$, with $|\psi'\rangle=\partial_x|\psi\rangle$, must be continuous there. This gives
\begin{equation}
\label{eq:bc}
\begin{aligned}
x=-d_x/2:&\quad
\begin{cases}
|\psi^{\text{I}}\rangle=|\psi^{\text{II}}\rangle\\
\hat{A}^{\text{I}}\partial_x|\psi^{\text{I}}\rangle=\hat{A}^{\text{II}}\partial_x|\psi^{\text{II}}\rangle+\tfrac{i}{2}\hat{B}^{\text{II}}|\psi^{\text{II}}\rangle
\end{cases},\\[4pt]
x=d_x/2:&\quad
\begin{cases}
|\psi^{\text{II}}\rangle=|\psi^{\text{III}}\rangle\\
\hat{A}^{\text{II}}\partial_x|\psi^{\text{II}}\rangle+\tfrac{i}{2}\hat{B}^{\text{II}}|\psi^{\text{II}}\rangle=\hat{A}^{\text{III}}\partial_x|\psi^{\text{III}}\rangle
\end{cases},
\end{aligned}
\end{equation}
where $\tfrac{i}{2}\hat{B}^{\text{II}}=i\alpha_R\hat{S}_y$ for the InAs nanowire and $\tfrac{i}{2}\hat{B}^{\text{II}}=\tfrac{i\alpha_{\text{OR}}}{2}(\cos\gamma\hat{L}_y-\sin\gamma\hat{L}_z)$ for the multiorbital one.

\begin{figure}
    \centering
    \includegraphics[width=0.6\linewidth]{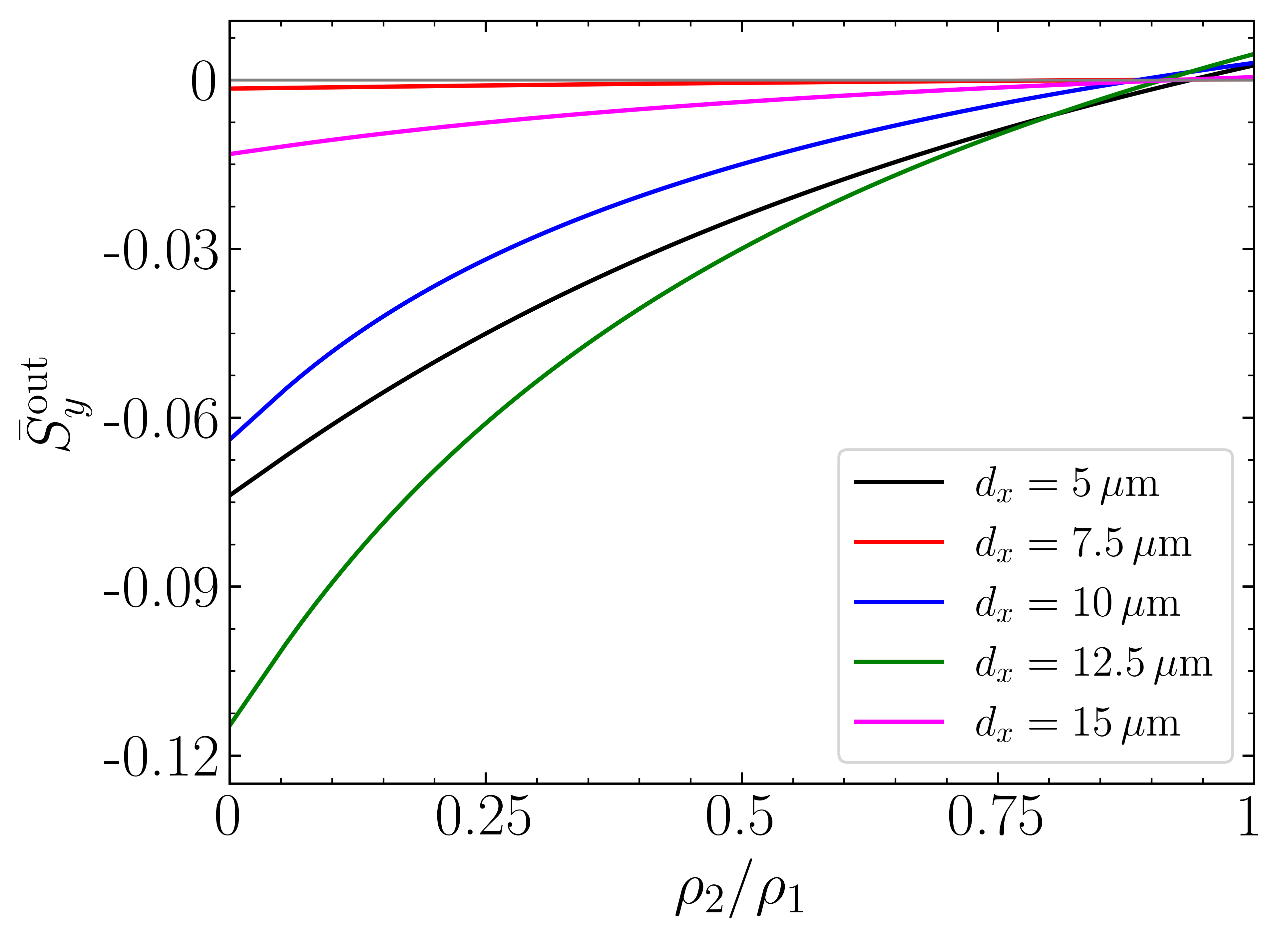}
    \caption{Average outgoing spin component $\bar{S}_y^{\text{out}}$ for a two-subband InAs nanowire as a function of the subband weight $\rho_2/\rho_1$ of the injecting lead, which is varied through a crystal-field term $\delta^{\text{I}}\hat{\tau}_z$ in lead I, for different wire lengths $d_x$. In all cases, the Fermi energy lies $2.5$~meV above the highest Kramers doublet of the nanowire at $k_x=0$. The model parameters are $a=25\,\epsilon_0/k_0^2$, $\alpha_R=2.6\,\epsilon_0/k_0$, $\beta_R=0.10\,k_0$, and $\Delta E=4.2\,\epsilon_0$, with $\epsilon_0=0.38$~meV and $k_0=0.1$~nm$^{-1}$. The metallic leads have a parabolic dispersion $a^{\text{I},\text{III}}k_x^2$ with $a^{\text{I},\text{III}}=30\,\epsilon_0/k_0^2$, and the bottom of their band lies $0.53$~meV below the lowest Kramers doublet of the nanowire.}
    \label{fig:S1}
\end{figure}

\begin{figure}
    \centering
    \includegraphics[width=0.6\linewidth]{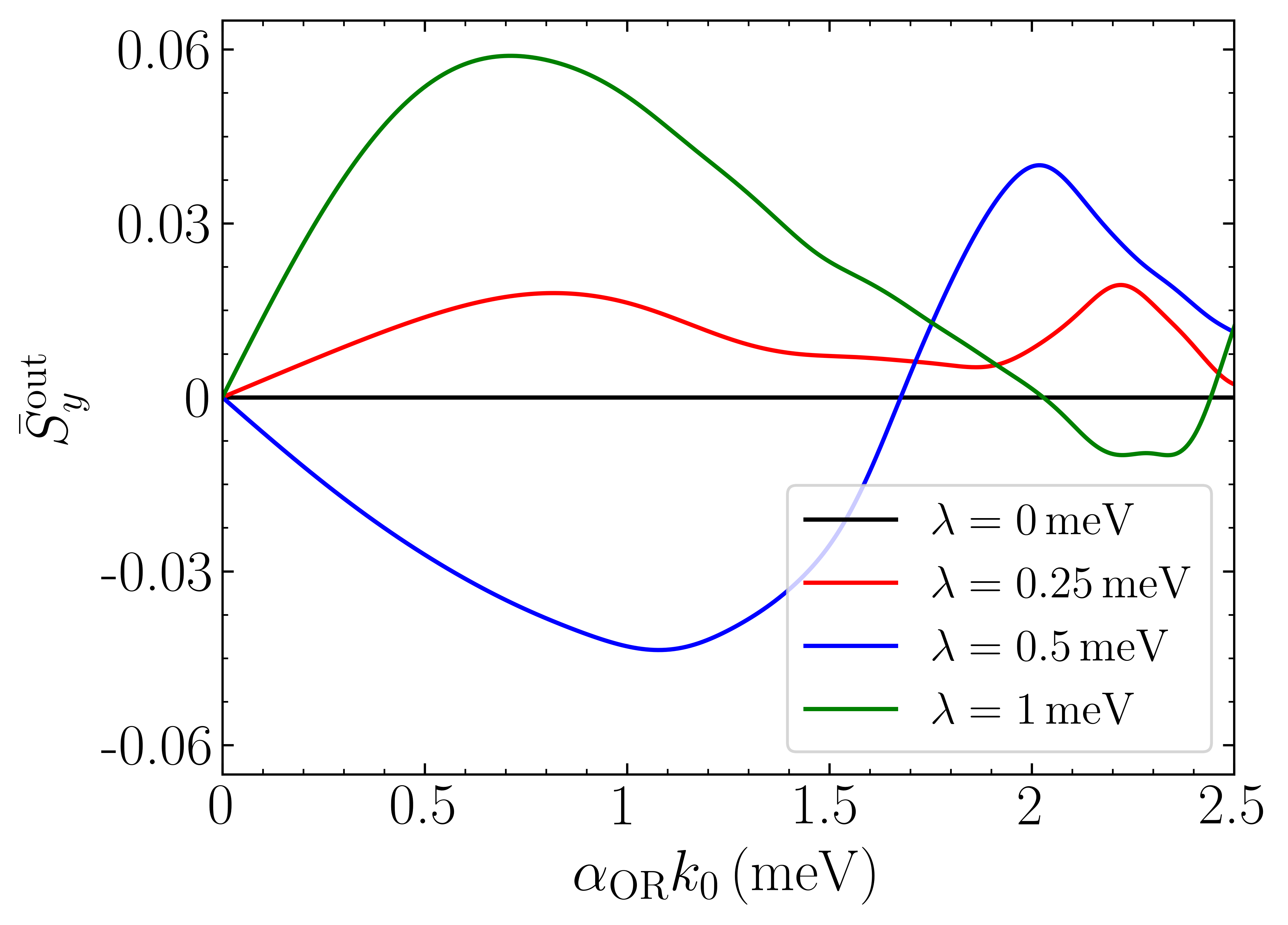}
    \caption{Average outgoing spin component $\bar{S}_y^{\text{out}}$ for a multiorbital nanowire as a function of the orbital Rashba coupling $\alpha_{\text{OR}}$, for selected values of the atomic spin-orbit coupling $\lambda$. The nanowire is modeled by the Hamiltonian $\mathcal{H}=\sum_{\alpha}\left(a_\alpha k_x^2+\delta_\alpha\right)\hat{L}_\alpha^2+\alpha_{\text{OR}}k_x\hat{L}_y+\lambda\hat{\mathbf{L}}\cdot\hat{\mathbf{S}}$, where $\hat{L}_\alpha$ ($\alpha=x,y,z$) are the $L=1$ angular momentum operators, $a_\alpha=0.15\,\epsilon_0/k_0^2$, and the crystal-field parameters are $\delta_x=\delta_y=-0.2\,\epsilon_0$ and $\delta_z=0.3\,\epsilon_0$, with $\epsilon_0=38$~meV and $k_0=1$~nm$^{-1}$. The two metallic leads host three parabolic bands with the same mass coefficients as the nanowire. In the source lead (region I), from which an unpolarized spin mixture is injected, we also include the crystal-field splittings $\delta_x^{\text{I}}=0.4\,\epsilon_0$, $\delta_y^{\text{I}}=-0.5\,\epsilon_0$, and $\delta_z^{\text{I}}=-0.7\,\epsilon_0$. The nanowire length is $d_x=1~\mu$m, and the Fermi energy lies about $20$~meV above the highest Kramers doublet of the nanowire.}
    \label{fig:S2}
\end{figure}

Denoting by $|\psi^{\text{A}}_{m,s}(x)\rangle$ the resulting state, we compute the local mean spin as
\begin{equation}
S_i^{m,s}(x)=\frac{\langle\psi^{\text{A}}_{m,s}(x)|\hat{S}_i|\psi^{\text{A}}_{m,s}(x)\rangle}{\langle\psi^{\text{A}}_{m,s}(x)|\psi^{\text{A}}_{m,s}(x)\rangle},\qquad i=x,y,z,
\end{equation}
where $\hat{S}_i=\hat{\sigma}_i/2$, and $\text{A}=\text{I}$ for $x\le-d_x/2$, $\text{A}=\text{II}$ for $-d_x/2<x<d_x/2$, and $\text{A}=\text{III}$ for $x\ge d_x/2$. We then average over the subband (orbital) channels. Since the spin-orbit-free leads possess an internal pseudospin (subband or orbital) degree of freedom, we weight the incoming subband (orbital) states according to the pseudospin content of the injected electrons, i.e., by the carrier densities of a one-dimensional free-electron gas, $n_m=2k_m^{\text{I}}/\pi$. The weights are thus $\rho_m=n_m/\sum_{m'}n_{m'}=k_m^{\text{I}}/\sum_{m'}k_{m'}^{\text{I}}$, and
\begin{equation}
S_i^{\uparrow}(x)=\sum_{m=1}^{N/2}\rho_m S_i^{m,\uparrow}(x),\qquad
S_i^{\downarrow}(x)=\sum_{m=1}^{N/2}\rho_m S_i^{m,\downarrow}(x),\qquad
\bar{S}_i(x)=\frac{S_i^{\uparrow}(x)+S_i^{\downarrow}(x)}{2}.
\end{equation}
The three panels of Fig.~\ref{figure3}(b) show $S_y^{\uparrow}(x)$, $S_y^{\downarrow}(x)$, and $\bar{S}_y(x)$ for the InAs nanowire with $\rho_1=\rho_2=1/2$, having the two subbands of lead I the same wavevector. All channels of region III also share the same wavevector, so these quantities are constant for $x\ge d_x/2$, where they give the outgoing values: $\bar{S}_i^{\text{out}}=\bar{S}_i(x\ge d_x/2)$. Figures~\ref{fig:S1} and \ref{fig:S2} show examples of the average outgoing spin component $\bar{S}_y^{\text{out}}$ for the InAs and the multiorbital nanowire, respectively.

\section*{Scattering matrix approach}

We now apply the Landauer approach
to compute the charge and spin conductances of the heterostructure introduced in the previous section. To this end, we normalize the propagating modes to unit flux,
\begin{equation}
|\psi_n^{\text{A},\pm}(x)\rangle=\frac{1}{\sqrt{|v_n^{\text{A},\pm}|}}\exp(ik_n^{\text{A},\pm}x)|u_n^{\text{A},\pm}\rangle,\qquad n=1,\ldots,N,
\end{equation}
where $v_n^{\text{A},\pm}=\langle u_n^{\text{A},\pm}|(2\hat{A}^{\text{A}}k_n^{\text{A},\pm}+\hat{B}^{\text{A}})|u_n^{\text{A},\pm}\rangle$ is the group velocity, positive for right-moving and negative for left-moving modes. In the leads, we again use the orthonormal product modes $|\chi_s\rangle\otimes|\phi_m\rangle$. A generic state in regions I and III is then
\begin{equation}
\begin{cases}
|\psi^{\text{I}}(x)\rangle=\sum_n\left[\alpha_{n}^{\text{I}}|\psi_{n}^{\text{I},+}(x)\rangle+\beta_{n}^{\text{I}}|\psi_{n}^{\text{I},-}(x)\rangle\right]\\
|\psi^{\text{III}}(x)\rangle=\sum_n\left[\alpha_{n}^{\text{III}}|\psi_{n}^{\text{III},-}(x)\rangle+\beta_{n}^{\text{III}}|\psi_{n}^{\text{III},+}(x)\rangle\right]
\end{cases},
\end{equation}
where $\alpha_n^{\text{A}}$ and $\beta_n^{\text{A}}$ are the amplitudes of the incoming and outgoing modes, respectively. The scattering matrix $S$ relates the vector of outgoing amplitudes, $\boldsymbol{\beta}=(\beta_{1}^{\text{I}}\:\dots\:\beta_{N}^{\text{I}}\:\beta_{1}^{\text{III}}\:\dots\:\beta_{N}^{\text{III}})^{T}$, to that of incoming amplitudes, $\boldsymbol{\alpha}=(\alpha_{1}^{\text{I}}\:\dots\:\alpha_{N}^{\text{I}}\:\alpha_{1}^{\text{III}}\:\dots\:\alpha_{N}^{\text{III}})^{T}$:
\begin{equation}
\boldsymbol{\beta}=S\boldsymbol{\alpha},\qquad S=\left(\begin{array}{cc}
r & \overline{t}\\
t & \overline{r}
\end{array}\right),
\end{equation}
where $r$ and $t$ ($\overline{r}$ and $\overline{t}$) are the $N\times N$ reflection and transmission matrices for electrons incoming from the left (right) lead. Since all modes carry unit flux, current conservation makes $S$ unitary.

We obtain $S$ by solving the scattering problem separately at the left ($L$) interface, between regions I and II, and at the right ($R$) one, between regions II and III, and then composing the two results. At each interface, we impose the boundary conditions discussed in the previous section (see Eq.~\ref{eq:bc}): since the probability current is $J=2\,\mathrm{Im}\langle\psi|\hat{A}|\psi'\rangle+\langle\psi|\hat{B}|\psi\rangle=2\,\mathrm{Im}\langle\psi|\tilde{\psi}\rangle$, the continuity of $|\psi\rangle$ and $|\tilde{\psi}\rangle$ ensures its conservation.

At the left interface, here placed at $x=0$, each right-moving mode of region I, $|\psi_n^{\text{I},+}\rangle$, is partly reflected back into region I and partly transmitted into the wire. The corresponding scattering state is
\begin{equation}
|\psi_{n}^{\text{I}}(x)\rangle=|\psi_{n}^{\text{I},+}(x)\rangle+\sum_{m}r_{m,n}^{L}|\psi_{m}^{\text{I},-}(x)\rangle
\end{equation}
in region I and
\begin{equation}
|\psi_{n}^{\text{II}}(x)\rangle=\sum_{m}t_{m,n}^{L}|\psi_{m}^{\text{II},+}(x)\rangle
\end{equation}
in region II. Imposing the continuity of $|\psi\rangle$ and $|\tilde{\psi}\rangle$ at $x=0$ for $n=1,\ldots,N$ gives the reflection and transmission matrices $r^L$ and $t^L$, with elements $r^L_{m,n}$ and $t^L_{m,n}$.

Similarly, each left-moving mode of the wire, $|\psi_n^{\text{II},-}\rangle$, is partly reflected back into the wire and partly transmitted into region I:
\begin{equation}
|\xi_{n}^{\text{II}}(x)\rangle=|\psi_{n}^{\text{II},-}(x)\rangle+\sum_{m}\overline{r}_{m,n}^{L}|\psi_{m}^{\text{II},+}(x)\rangle,
\end{equation}
\begin{equation}
|\xi_{n}^{\text{I}}(x)\rangle=\sum_{m}\overline{t}_{m,n}^{L}|\psi_{m}^{\text{I},-}(x)\rangle.
\end{equation}
The same boundary conditions give $\overline{r}^L$ and $\overline{t}^L$, which complete the scattering matrix of the left interface:
\begin{equation}
S^{L}=\left(\begin{array}{cc}
r^{L} & \overline{t}^{L}\\
t^{L} & \overline{r}^{L}
\end{array}\right).
\end{equation}

We apply the same procedure to the right interface, whose scattering matrix $S^{R}$ is computed as if this interface were also placed at $x=0$. The wire length $d_x$ thus appears only in the propagation factors of the wire modes, $f_{\pm}=\mathrm{diag}[\exp(ik_{n}^{\text{II},\pm}d_x)]_{n=1,\ldots,N}$, through which the composition of $S^L$ and $S^R$ gives the scattering matrix of the heterostructure:
\begin{equation}
\begin{cases}
r=r^{L}+\overline{t}^{L}\Omega^{-1}r^{R}f_{+}t^{L}\\
\overline{t}=\overline{t}^{L}\Omega^{-1}\overline{t}^{R}\\
t=t^{R}f_{+}t^{L}+t^{R}f_{+}\overline{r}^{L}\Omega^{-1}r^{R}f_{+}t^{L}\\
\overline{r}=t^{R}f_{+}\overline{r}^{L}\Omega^{-1}\overline{t}^{R}+\overline{r}^{R}
\end{cases},
\end{equation}
with $\Omega=f_{-}-r^{R}f_{+}\overline{r}^{L}$.

Finally, in linear response and at zero temperature, the charge and spin conductances for unpolarized injection from the left lead, i.e., with all incoming modes equally populated, follow from the transmission matrix $t$ through the Landauer formula:
\begin{equation}
G=\frac{e^2}{h}\,\mathrm{Tr}\left(tt^\dagger\right),\qquad G_i=\frac{e^2}{h}\,\mathrm{Tr}\left[tt^\dagger\left(\mathbb{I}_{N/2} \otimes\hat{\sigma}_i\right)\right],
\end{equation}
where $i=x,y,z$ and $\mathbb{I}_{N/2}$ is the identity in the subband or orbital space.

\section*{Supplemental figures}

\begin{figure}
    \includegraphics[width=0.9\textwidth]{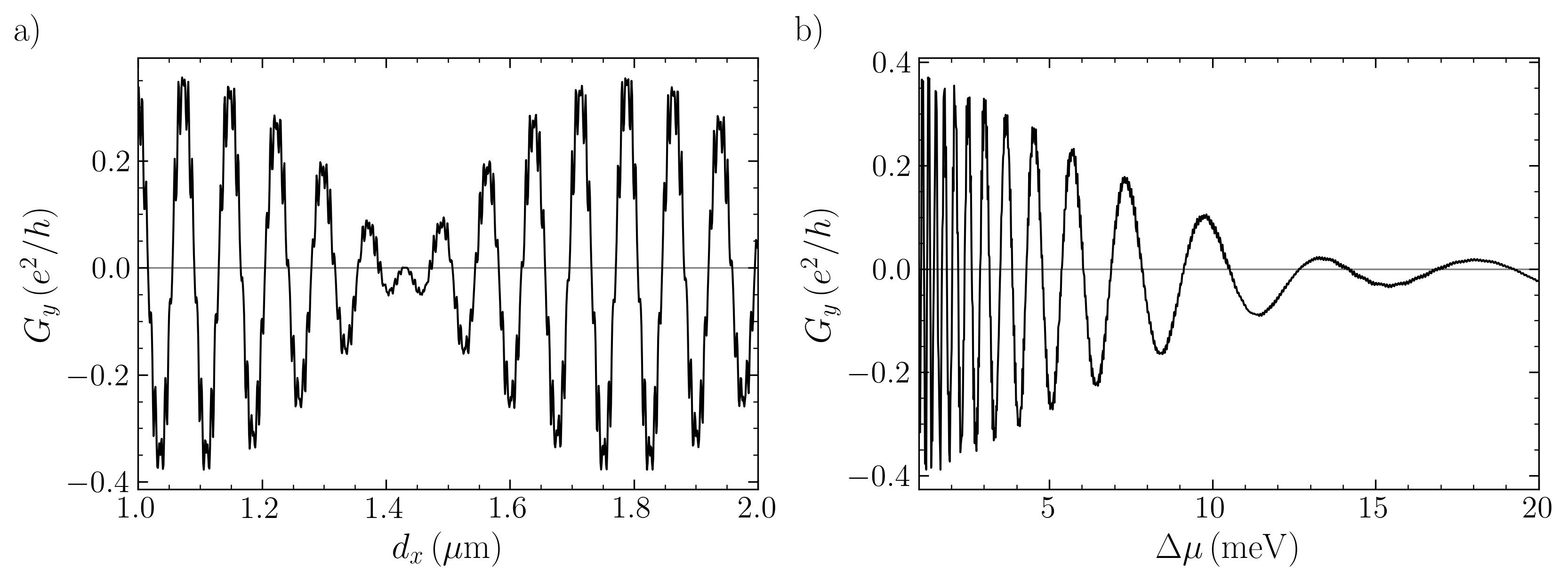}
    \centering
    \caption{Spin conductance $G_y$ of a quasi-1D wire described by the $L=1$ model, with $\lambda=1$~meV and $\alpha_{\mathrm{OR}}k_0=2$~meV, computed within the scattering-matrix formalism; $\Delta\mu$ denotes the energy separation between the chemical potential $\mu$ and the highest Kramers doublet of the wire at $\Gamma$. (a) $G_y$ as a function of the wire length $d_x$, for $\Delta\mu=2.5$~meV. (b) $G_y$ as a function of $\Delta\mu$, for $d_x=1~\mu$m. All other parameters are as in the main text. In both cases, $G_y$ oscillates and changes sign.}
    \label{fig:R4}
\end{figure}

\begin{figure}
    \includegraphics[width=0.6\textwidth]{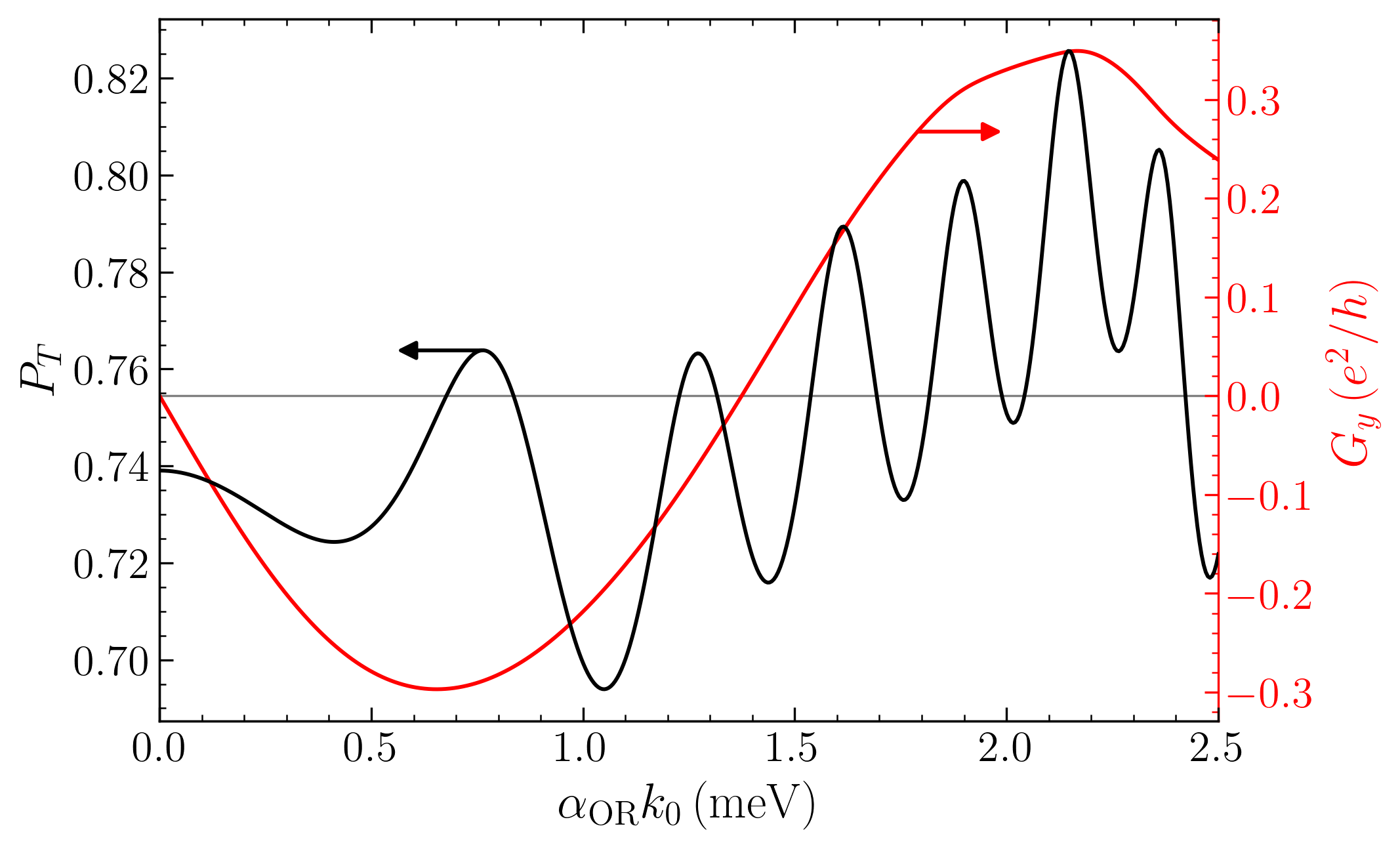}
    \centering
    \caption{Single-electron transmission probability $P_T$ (black, left axis) and spin conductance $G_y$ (red, right axis) as a function of the orbital Rashba coupling $\alpha_{\mathrm{OR}}k_0$, for a quasi-1D wire described by the $L=1$ model. Here, $\lambda=1$~meV, $d_x=1~\mu$m, and the chemical potential is $2.5$~meV above the highest Kramers doublet of the wire at $\Gamma$; all other parameters are as in the main text.}
    \label{fig:R5}
\end{figure}

\begin{figure}
    \centering
    \includegraphics[width=0.6\linewidth]{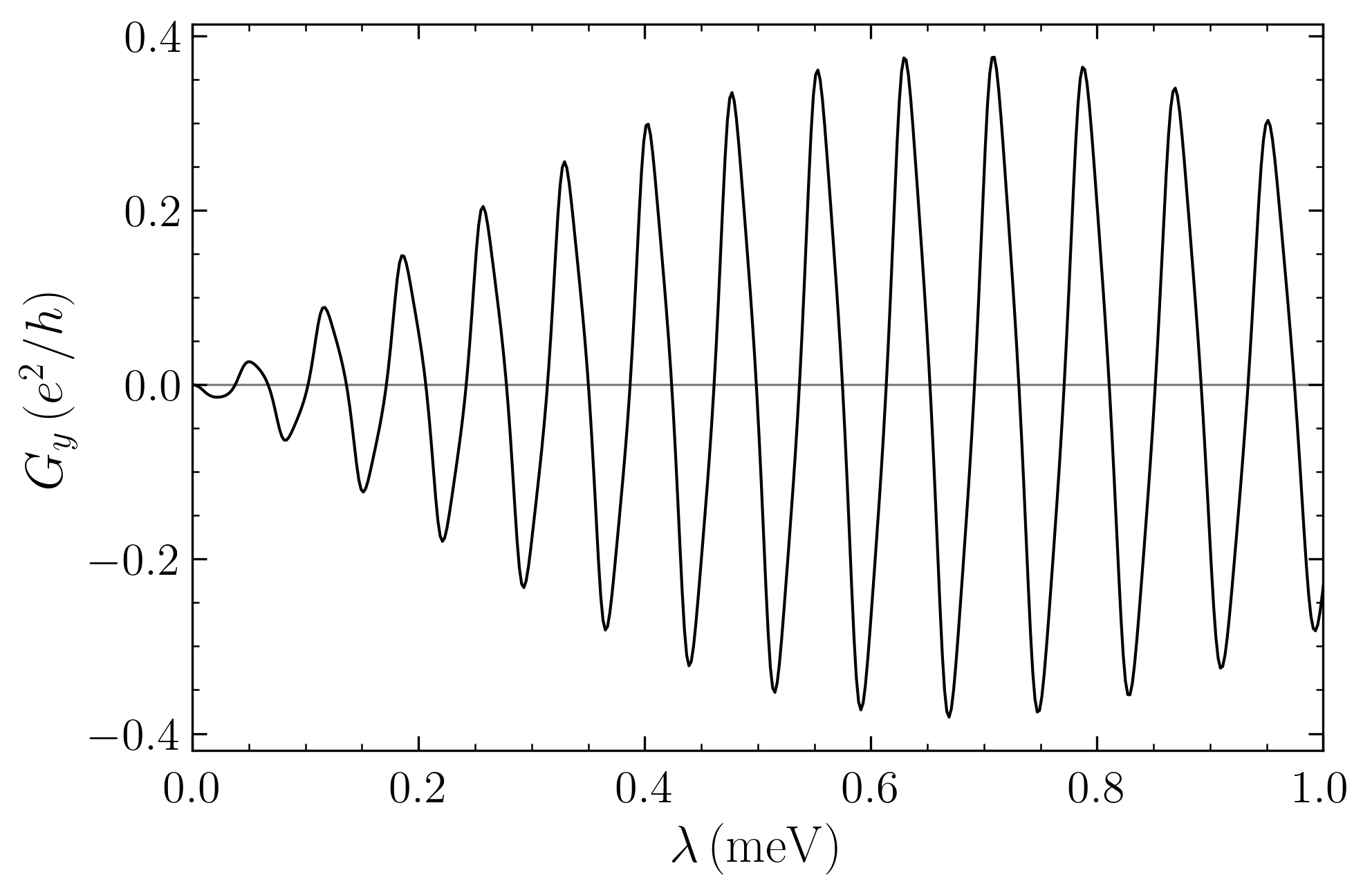}
    \caption{Spin conductance $G_y$ as a function of the atomic spin--orbit coupling $\lambda$ for a quasi-1D multiorbital ($L=1$) wire, computed within the scattering-matrix formalism for unpolarized injection. Here, $\alpha_{\mathrm{OR}}k_0=1$~meV, $d_x=1~\mu$m, and the chemical potential is $2.5$~meV above the highest Kramers doublet at $\Gamma$ within the wire; all other parameters are as in the main text.}
    \label{fig:R8}
\end{figure}

\end{document}